\documentclass[aps,twocolumn]{revtex4-1}

\usepackage{graphicx}
\usepackage{multirow}
\usepackage{xcolor}
\usepackage{units}
\usepackage{mathrsfs}
\usepackage{mathptmx, textcomp}
\usepackage{amssymb, amsmath}

\usepackage[]{lineno}

\usepackage{caption}
\DeclareCaptionLabelSeparator{vline}{ | }
\makeatletter
\def\justified{
  \let\\\@normalcr
  \@rightskip\z@skip \rightskip\@rightskip
  \leftskip\z@skip
  \parindent 0em\relax
  \setlength{\parfillskip}{0pt plus 1fil}}
\DeclareCaptionJustification{justified}{\justified}

\DeclareCaptionType{extfig}
 \newcommand{\bs}{\boldsymbol}
\newcommand{\Er}{\ensuremath{^{166}}{\rm Er} }

\newcommand{\br}{\ensuremath{\bs{r }}}

\newcommand{\as}{\ensuremath{a_{\rm s}}}
\newcommand{\add}{\ensuremath{a_{\rm dd}}}
\newcommand{\tho}{t_{\rm h}}
\newcommand{\ttof}{t_{\rm f}}
\newcommand{\ntof}{n_{\rm tof}}
\newcommand{\um}{\mu{\rm m}}
\newcommand{\kr}{k_{\rm rot}}
\newcommand{\krot}{k_{\rm rot}}

\newcommand{\ar}{a_{\rm s}^{*}}
\newcommand{\af}{a_{\rm s}^{\rm f}}
\newcommand{\asi}{a_{\rm s}^{\rm i}}

\newcommand{\crot}{\ensuremath{\mathcal{C}}}
\newcommand{\epsdd}{\epsilon_{\rm dd}}

\newcommand{\rvec}{\mathbf r}
\newcommand{\brho}{\bs \rho}
\newcommand{\tx}{\ensuremath{{\tilde x }}}
\newcommand{\tz}{\ensuremath{{\tilde z }}}
\newcommand{\lp}{\lambda_{\perp}}

\renewcommand{\Im}{\operatorname{Im}}

\begin{document}

\title{Observation of Roton Mode Population in a Dipolar Quantum Gas}

\author{L. Chomaz$^1$, R. M. W. van Bijnen$^{2}$, D. Petter$^{1}$, G. Faraoni$^{1,3}$, S. Baier$^{1}$,  J. H. Becher$^{1}$, M. J. Mark$^{1,2}$, F. W\"achtler$^4$, L. Santos$^4$, F. Ferlaino$^{1,2,*}$}

\affiliation{%
 $^{1}$Institut f\"ur Experimentalphysik,Universit\"at Innsbruck, Technikerstra{\ss}e 25, 6020 Innsbruck, Austria\\
$^{2}$Institut f\"ur Quantenoptik und Quanteninformation,\"Osterreichische Akademie der Wissenschaften, 6020 Innsbruck, Austria\\
$^{3}$Dipartimento di Fisica e Astronomia, Universit\`a di Firenze, Via Sansone 1, 50019 Sesto Fiorentino, Italy\\
$^{4}$Institut f\"ur Theoretische Physik, Leibniz Universit\"at Hannover, Appelstr. 2, 30167 Hannover, Germany}%

\maketitle



{\bf \noindent The concept of a roton, a special kind of elementary excitation, forming a minimum of energy at finite momentum, has been essential to understand the properties of superfluid $^4$He~\cite{Landau41}. In quantum liquids, rotons arise from the strong interparticle interactions, whose microscopic description remains debated~\cite{Griffin:1993}.  In the realm of highly-controllable quantum gases, a roton mode has been predicted to emerge due to magnetic dipole-dipole interactions despite of their weakly-interacting character~\cite{Santos:2003}. This prospect has raised considerable interest \cite{Ronen:2007,Bohn:2009,Parker:2009,Martin:2012,Blakie:2012,Lasinio:2013,Wilson:2010,Natu:2014,Pitaevskii:2016}; yet roton modes in dipolar quantum gases have remained elusive to observations. Here we report experimental and theoretical studies of the momentum distribution in Bose-Einstein condensates of highly-magnetic erbium atoms, revealing the existence of the long-sought roton mode. 
By quenching the interactions, we observe the roton appearance of peaks at well-defined momentum. The roton momentum follows the predicted geometrical scaling with the inverse of the confinement length along the magnetisation axis. From the growth of the roton population, we probe the roton softening of the excitation spectrum in time and extract the corresponding imaginary roton gap. Our results provide a further step in the quest towards supersolidity in dipolar quantum gases~\cite{Boninsegni:2012}.}

Quantum properties of matter continuously challenge our intuition, especially when many-body effects emerge at a macroscopic scale. 
In this regard, the phenomenon of superfluidity is a paradigmatic case, which  continues to reveal fascinating facets since its discovery in the late 1930s~\cite{Griffin:1993,Pitaevskii:2016}.
A major breakthrough in understanding superfluidity thrived on the concept of {\em quasiparticles}, introduced by Landau in 1941~\cite{Landau41}. Quasiparticles are elementary excitations of momentum $k$, whose energies $\epsilon$ define the dispersion~(energy-momentum) relation $\epsilon(k)$. 

To explain the special thermodynamic properties of superfluid $^{4}\rm{He}$,
Landau postulated the existence of two types of low-energy quasiparticles: phonons, referring to low-$k$ acoustic waves, and rotons, gapped excitations at finite $k$ initially interpreted as elementary vortices. The dispersion relation continuously evolves from linear at low $k$ (phonons) to  
parabolic-like with a minimum (roton) at a finite $k=\kr$. Neutron scattering experiments confirmed Landau`s remarkable intuition~\cite{Henshaw61}. In liquid $^4$He, $\kr$ 
scales as the inverse of the interatomic distance. This manifests a tendency of the system to establish a local order, which is  driven by the strong correlations among the atoms~\cite{Griffin:1993}. 

In the realm of low-temperature quantum physics, ultra-cold quantum gases realise the other extreme limit for which the interparticle  interactions - and correlations - are typically weak, meaning that classically their range of action is much smaller than the mean interparticle distance. Because of this diluteness, roton excitations are absent in ordinary quantum gases, i.\,e.\,in Bose-Einstein condensates (BECs) with contact (short-range) interactions~\cite{Pitaevskii:2016}. 
However, about 15 years ago, seminal theoretical works predicted the existence of a
roton minimum both in BECs with magnetic dipole-dipole interactions (DDIs)~\cite{Santos:2003} and in BECs irradiated by off-resonant laser light~\cite{Odell2003}. Following the lines of the latter proposal, a roton softening  has been recently observed in BECs coupled to an optical cavity~\cite{Mottl12}. Here, the excitation arises from the infinite-range photon-mediated interactions and the inverse of the laser wavelength sets the value of $\kr$. Additionally, roton-like softening has been created in spin-orbit-coupled BECs~\cite{Khamehchi:2014,Ji2015} and quantum gases in shaken optical lattices~\cite{LiChung15} by engineering the single-particle dispersion relation.

Our work focuses on dipolar BECs (dBECs). As in superfluid $^4$He, the roton spectrum in such systems is a genuine consequence of the underlying interactions among the particles.
However, in contrast to helium,  the emergence of a minimum at finite momentum does not require strong inter-particle interactions. It instead exists in the weakly-interacting regime and originates from the peculiar anisotropic and long-range character of the DDI in real and momentum space~\cite{Santos:2003,Ronen:2007,Bohn:2009,Parker:2009,Martin:2012,Blakie:2012,Lasinio:2013,Wilson:2010,Natu:2014,Pitaevskii:2016}.  
Despite the maturity achieved in the theoretical understanding, the observation of dipolar roton modes has remained so far an elusive goal.
For a long time, the only dBEC available in experiments consisted of chromium atoms~\cite{Griesmaier:2005}, for which the achievable dipolar character is hardly sufficient to support a roton mode. With the advent of the more magnetic lanthanide atoms~\cite{Lu:2011,Aikawa:2012}, a broader range of dipolar parameters became available, opening the way to access the regime of dominant DDI. 
In this regime, novel exciting many-body phenomena have been recently observed, as the formation of droplet states stabilised by quantum fluctuations~\cite{Kadau:2016,Igor:2016,Chomaz:2016}, which may become self-bound~\cite{Schmitt:2016}. 
Lanthanide dBECs hence open new roads toward the long-sought observation of roton modes.

\begin{figure}
  \centering
\includegraphics{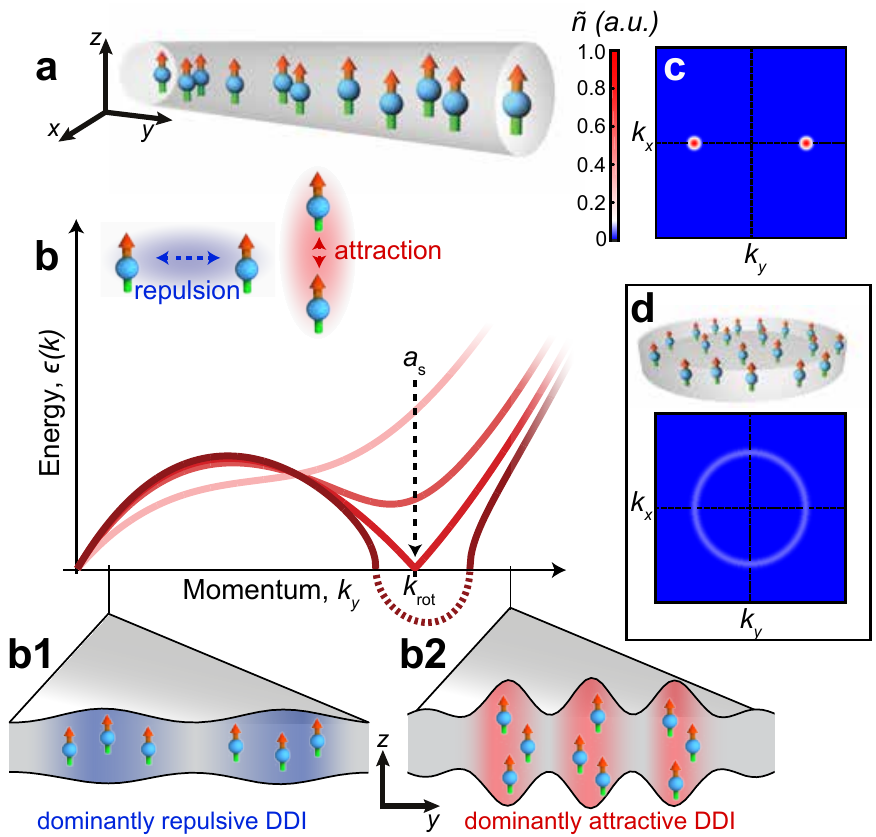} 
\caption{\label{fig1}
{\bf Roton mode in a dBEC}. 
{\bf a}, axially elongated geometry with dipoles oriented transversely. 
{\bf b}, real (solid lines) and imaginary (norm of the dotted line) parts of the dispersion relation of a dBEC in the geometry {\bf a}, showing the emergence of a roton minimum for decreasing $\as$ (dashed arrow). The DDI changes from repulsive (blue) to attractive (red) depending on the dipole alignment (inset). {\bf b1, b2}, the dipole alignment (color code as in inset) associated to small- ({\bf b}1) and large-$k_y$ ({\bf b}2)  density modulations.  
 {\bf c, d}, distributions on the $k_x\,k_y$-plane associated with the roton population in cigar {\bf a} or pancake {\bf d} geometries with an identical roton population and colorscale.
}
\end{figure}

Prior to this work, dipolar rotons have been mostly connected to pancake-like geometries~\cite{Santos:2003,Ronen:2007,Bohn:2009,Parker:2009,Blakie:2012,Lasinio:2013,Wilson:2010,Natu:2014,Pitaevskii:2016}.   
Here, we extend the study of roton physics to the case of a cigar-like geometry  with trap elongation along only one direction ($y$) transverse to the magnetisation axis ($z$) (Fig.\,1a).  The anisotropic character of the DDI (Fig.\,1b, inset) together with the tighter confinement along $z$  is responsible for the rotonization of the excitation spectrum along $y$ (Fig.\,1b). To illustrate this phenomenon, we consider an infinite cigar-shaped dBEC and focus on its axial elementary excitations, of momentum $k_y$. These excitations correspond in real space to a density modulation along $y$ of wavelength $2\pi/k_y$. For low $k_y$, the atoms sit mainly side-by-side and the repulsive nature of the DDI prevails, stiffening the phononic part of the dispersion relation (Fig.\,1b1). In contrast, for $k_y\ell_z \gtrsim 1$, $\ell_z$ being the characteristic $z$ confinement length, the excitation favours head-to-tail alignments and the DDI contribution to $\epsilon(k_y)$ eventually changes sign~\cite{Santos:2003} (Fig.\,1b2). The resulting softening of $\epsilon(k_y)$ is counterbalanced by the contributions of the repulsive contact interaction, and of the kinetic energy, which ultimately dominates at very large $k_y$. For strong enough DDI, this competition gives rise to a roton minimum in $\epsilon(k_y)$, occurring at a momentum $k_y =\krot$ 
set by the geometrical scaling $\krot \sim 1/\ell_z$~(see below and e.g.\,\cite{Santos:2003,Blakie:2012,Lasinio:2013}).

Similar to the helium case, the roton energy gap, $\Delta = \epsilon(\kr)$, depends on the density and on the strength of the interactions. In ultracold gases, both quantities can be controlled. In particular,  the scattering length $\as$, setting the strength of the contact interaction, can be tuned using Feshbach resonances~\cite{Pitaevskii:2016}. As $\as$ is reduced, $\Delta$ decreases, vanishes, and eventually becomes imaginary (Fig.\,1b). 
In the latter case, the system undergoes a roton  instability  and the population at $k_y=0$ is transferred in $\pm \kr$ at an exponential rate~\cite{Bohn:2009,Parker:2009}.
The population of the roton mode is then readily visible in the momentum distribution of the gas (Fig.\,1c-d). In the extensively-studied pancake geometries, the roton population in $k$-space spreads over a ring of radius $k=\kr$ because of the radial symmetry of the confinement (Fig.\,1d). Such a spread can be  avoided using a cigar geometry. Here, the roton population focuses in two prominent peaks at $k_y=\pm \kr$, enhancing the visibility of the effect (Fig.\,1c). 

We explore the above-described physics using strongly magnetic $\Er$ atoms.
The experiment starts with a stable dBEC
in a cigar-shaped harmonic trap of frequencies $\nu_{x,y,z}$, elongated along the $y$ axis. The trap aspect ratio, $\lambda=\nu_z/\nu_y$, can be tuned from about 4 to 30, corresponding to $\nu_z$ ranging from 150\,Hz to 800\,Hz, whereas $\nu_y$ and $\nu_z/\nu_x$ are kept constant at about $35\,$Hz and $ 1.6$, respectively~(Methods).  
An external homogeneous magnetic field, $B$, 
fixes the dipole orientation (magnetisation) with respect to the trap axes and sets the values of $\as$ through a magnetic Feshbach resonance, centered close to $B=0\,$G. In previous experiments, we precisely calibrated the $B$-to-$\as$ conversion for this resonance\,\cite{Chomaz:2016}. The BEC is prepared at $\asi=61\,a_0$ ($B=0.4\,$G) with transverse ($z$) magnetisation. The characteristic dipolar length
, defined as $\add=\mu_0\mu_m^2m/12\pi\hbar^2$, is $65.5\,a_0$, with $m$ the mass and $\mu_m$ the magnetic moment of the atoms, $\hbar=h/2\pi$ is the reduced Planck constant, $\mu_0$ the vacuum permeability and $a_0$ the Bohr radius. 

\begin{figure}
  \centering
\includegraphics{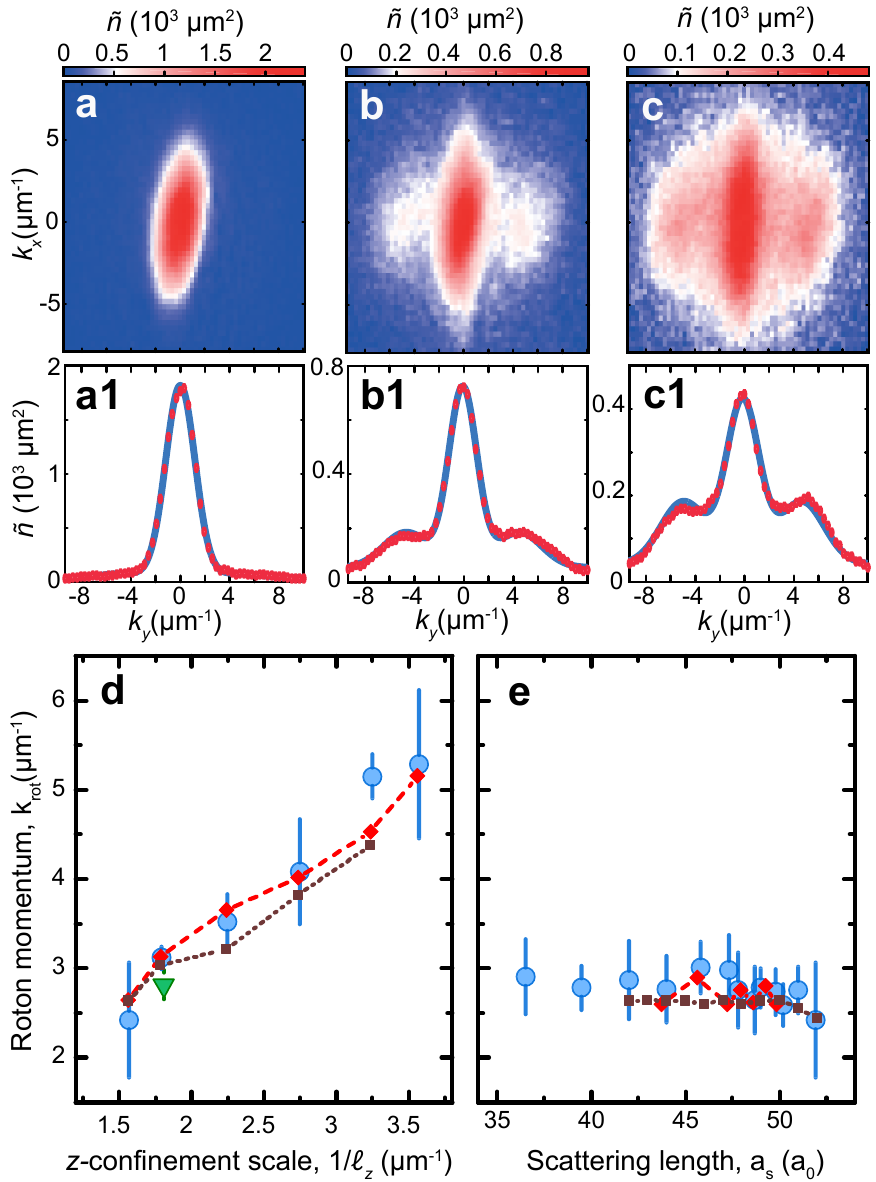} 
\caption{\label{fig:scaling}
{\bf Observed roton peaks and characteristic scalings}. {\bf a-c}, $\tilde{n}(k_x,k_y)$ obtained by averaging 15-to-25 absorption images for $(\nu_z,\lambda)=(456\,{\rm Hz},14.4)$, $\tho=3\,$ms and:  
{\bf a}, $\as=54a_0$, {\bf b}, $\as=44a_0$ and {\bf c}, $\as=37a_0$. 
{\bf a1-c1}, corresponding cuts at $k_x\approx 0$ (dots) and their fits to three-Gauss distributions (lines){, from which we extract $\kr$ and $A_*$.}  
{\bf d}, Measured $\kr$ depending on  $1/\ell_z$   at $\as \lesssim \ar$  for $\nu_y\approx 35\,$Hz (circles) and  $\nu_y=17(1)\,$Hz (triangle). {\bf e}, $\kr$ depending on $\as$ for $(\nu_z,\lambda)=(149\,{\rm Hz},4.3)$ (circles). 
{In (d-e), error bars show the 95\% confidence interval of the three-Gauss fits (see a1-c1).} 
The squares (diamonds) show predictions from the SSM (NS) in their range of validity (Methods). Lines are guides-to-the-eye.
} 
\end{figure}

To excite the roton mode, we quench $\as$ to a desired lower value, $\af$, and shortly hold the atoms in the trap for a time $\tho$. We measure that $\as$ converges to its set value with a characteristic time constant of $1\,{\rm ms}$ during $\tho$. We then release the atoms from the trap, change $\as$ back close to its initial value and let the cloud expand for $30\,$ms. We probe the momentum distribution $\tilde{n}(k_x,k_y)$ by performing standard resonant absorption imaging on the expanded cloud~(Methods). The measurement is then repeated at various values of $\as<\add$ in a fixed  trap geometry. The momentum distribution shows a striking behaviour (Fig.\,2). For large enough $\as$, $\tilde{n}(k_x,k_y)$ shows a single narrow peak with an inverted aspect ratio compared to the trapped gas, typical of a stable BEC \cite{ Pitaevskii:2016} (Fig.\,2a). 
We define the center of the distribution as the origin of $k$. 
In contrast, when further decreasing $\as$ below a critical value $\ar$, 
we observe a sudden appearance  of 
two symmetric finite-momentum peaks, of similar shape and located at $k_y=\pm \kr$~(Fig.\,2b-c). By repeating the experiment several times, we observe that the peaks consistently appear at the same locations, and they are visible in the averaged distributions.
To quantitatively investigate the peak structures, we fit a sum of three Gaussian distributions to the central cuts of the average $\tilde{n}(k_x,k_y)$ (Fig.\,2a1-c1). From the fit we extract the central momentum, $\kr$, and the amplitude, $A_{*}$, of the side peaks.   

A major fingerprint of the roton mode in dBECs is its geometrical nature, leading to a universal scaling $\kr \sim 1/\ell_z$  (see e.g.\,\cite{Santos:2003,Ronen:2007,Blakie:2012,Lasinio:2013}). In addition, the dependency of $\krot$ on $\as$ close to the instability is expected to be mild as $\kr$ remains mainly set by its geometrical nature \cite{Santos:2003,Blakie:2012}. 
We investigate both dependencies in the experiment. In a first set of experiments, we repeat the quench measurements for various trap parameters and extract $\krot$.  We clearly observe the expected geometrical scaling. 
$\krot$ shows a marked increase with $1/\ell_z =2\pi\sqrt{m\nu_z/h}$, matching well with a linear progression with a slope of 1.61(4) (Fig.\,2d). 
Note that no dependence of $\krot$ on $\nu_y$ is observed. In a second set of experiments, we fix the trap geometry and explore the dependence of $\krot$ on $\as$.
We observe that, within our experimental uncertainty, $\krot$ stays constant when decreasing $\as$ ~(Fig.\,2e). This behaviour contrasts the one expected for a phonon-driven modulation instability that exhibits a strong $\as$-dependence~\cite{Nguyen:2017}.

To gain a deeper understanding of the roton excitations in our system and its dynamical population, we develop both an analytical model and full numerical simulations and compare the findings with our  experimental data. 
Our analytical model starts by calculating the roton spectrum of the stationary BEC, generalizing the results of Ref.\,\cite{Santos:2003} to a non-radially symmetric configuration.
Since the roton wavelength is much smaller than the extension of our 3D BEC along $y$, this mode can be evaluated using a local density approximation in $y$. Hence, for our model, we consider a dBEC homogeneous along $y$, of axial density $n_0$, harmonically confined along $x$ and $z$. To analytically evaluate the roton spectrum, we approximate the BEC wavefunction using the Thomas-Fermi~(TF) approximation. 
For dominant DDI, $\epsdd=\add/\as\geq 1$, we find that $\epsilon(k_y)$ indeed rotonizes~(Methods). In the vicinity of the roton minimum and for $\epsdd \sim 1$, the computed dispersion acquires a gapped quadratic form similar to that of  helium rotons:
\begin{equation}
\epsilon(k_y)^2\simeq \Delta^2 + \frac{2\hbar^2\kr^2}{m}
\frac{\hbar^2}{2m}(k_y-\kr)^2.
\label{eq:homogeneous_spectrum}
\end{equation}
The roton momentum reads as $\kr = \sqrt{2m}(E_0^2-\Delta^2)^{1/4}/\hbar$, with $\Delta=\sqrt{E_0^2-E_I^2}$, $E_I=2gn_0 (\epsilon_{dd}-1)/3$, $g=4\pi\hbar^2 \as/m$, and $E_0^2=2 g\epsilon_{dd} n_0 \frac{\hbar^2}{2m}\left ( X^{-2} + Z ^{-2} \right )$. 
The TF radii verify $X^2, Z^2 \propto gn_0$ so that $E_0 \propto h\nu_z$, with a scale set by $\epsdd$ and $\nu_z/\nu_x$ but independent of $g n_0$~\cite{Odell2003}. 
Close to the instability ($\Delta\simeq 0$), the roton momentum thus follows a simple geometrical scaling, $\kr = \kappa/\ell_z$ with the 
geometrical factor $\kappa$ depending on $\nu_z/\nu_x$ alone. 
For completeness, we have also performed full 3D numerical calculations of the static Bogoliubov spectrum of a finite trapped dBEC, confirming the existence and scaling of the roton mode~(Supplementary Information).

\begin{figure*}
  \centering
\includegraphics{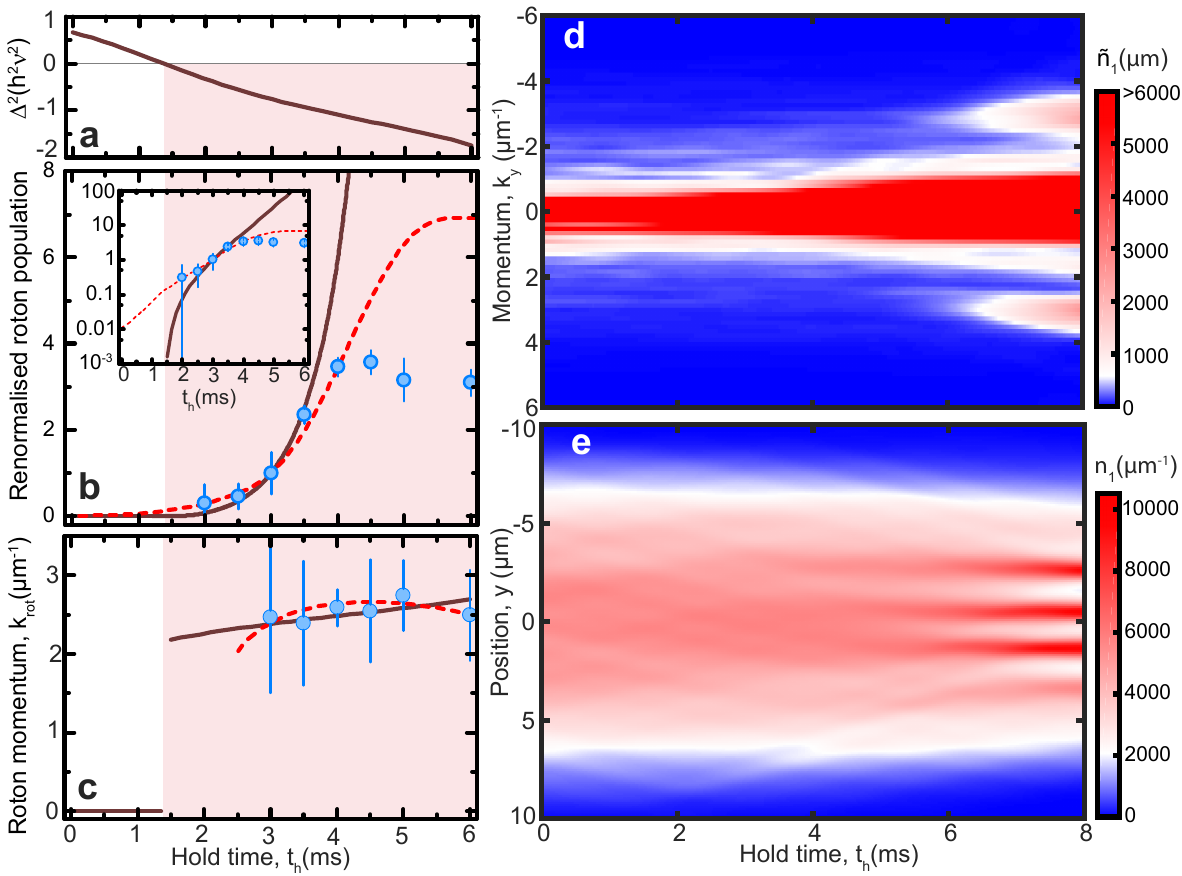} 
\caption{\label{fig3} 
{\bf Dynamics of the roton mode.} Results depending on $\tho$ after quenching to $\af=50\,a_0$ in $(\nu_z,\lambda)=(149\,{\rm Hz},4.3)$. {\bf a-c}, $\Delta^2$ at $y=0$,  amplitude and momentum of the roton peak, from the SSM (solid lines), the NS (dashed lines) and the experiments (circles) with their {95\% confidence interval from the three-Gauss fit (error bars)}. Shaded areas identify the dynamical destabilization.  For comparison, the amplitudes are renormalized to their $\tho=3\,$ms-values and the NS results are translated by -3.7\,ms. $\krot$ is reliably extracted from the experiments (NS) for $\tho\geq\,$3\,ms (2.5\,ms).   Inset, same plot in log scale.  {\bf d, e}, Integrated 1D density profile, in momentum $\tilde{n}_1(k_y)$ and real space $n_1(y)$, from one run of the NS.
} 
\end{figure*}

The above stationary description accounts for the existence of the roton mode in the cigar geometry used in experiments, and predicts the scaling of $\kr$ and $\Delta$ with the system parameters. However the quench of $\as$ introduces a dynamic, which is crucial to quantitatively reproduce the experimental observations. The reduction of $\as$ decreases the contact interaction and additionally induces a compression of the cloud. This  yields a dynamical modification of the local roton dispersion relation, and the roton may be destabilised during the evolution.
This dynamical destabilisation is well accounted for by a self-similar model (SSM) describing the  
evolution of the cloud shape after the quench \cite{Pitaevskii:2016}. 
In particular, we consider a 3D harmonic confinement, and, starting from the 
stationary TF profile at $\asi$, we evaluate the evolution of the cloud 
along the change of $\as$ assuming that the TF shape is maintained but with time-dependent TF radii~(Methods). 
We then estimate the
local~(along $y$) instantaneous roton spectrum $\epsilon(k_y,y,\tho)$ using a local density approximation, i.e. evaluating 
Eq.\,\eqref{eq:homogeneous_spectrum} with  the experimentally calibrated $\as(\tho)$, and the $n_0(y,\tho)$, $X(y,\tho)$, 
$Z(y,\tho)$ estimated from the 3D profile~(Methods). 
We indeed find that the roton gap decreases 
and eventually turns imaginary 
($\Delta(\tho)^2<0$)~(Fig.\,3a). When this occurs
the population at $k_y\approx \krot$ exponentially grows with an instantaneous local rate $2\Im[\epsilon(k_y,y,\tho)]/\hbar$, $\Im[.]$ indicating the imaginary part, giving rise to two symmetric side peaks in the axial momentum
distribution $\tilde{n}_1( k_y,\tho)\propto
\int\exp\left(2\int_0^{\tho}\Im[\epsilon(k_y,y,t)]dt\right)dy$ 
~(Fig.\,3b) (Methods).
The center of the peaks, $\pm\kr(\tho)$, also evolves with $\tho$ but quickly converges after a few ms to its final value, $\kr$~(Fig.\,3c). The measured $\kr$ and the calculated values from our parameter-free theory are in remarkable agreement (Fig.\,2d-e and 3c).

Our SSM quantitatively explains the experimental observations and provides us with a physical understanding. For completeness, we perform numerical simulations (NS) of the system dynamics. We calculate the time evolution of the generalized non-local Gross-Pitaevskii equation, which accounts also for
 quantum fluctuations, three-body loss processes, and finite temperature,  
not included in the SSM~(Supplementary Information).  
The first two contributions limit the peak density of the atomic cloud and stabilise the dBEC against collapse~\cite{Ronen:2007, Bohn:2009, Parker:2009, Igor:2016,Chomaz:2016}, whereas the latter term thermally seeds the initial roton-mode population. 
The NS confirm both the SSM results and the experimental observations. Indeed the calculations show that, few ms after the quench, the system develops roton peaks in momentum space (Fig.\,3d) and short-wavelength density modulations at the center of the BEC~(Fig.\,3e), showing the predicted roton confinement \cite{Lasinio:2013}. The extracted value of $\krot$ and its geometrical scaling are in very good quantitative agreement with both the experimental data and the SSM calculations (Fig.\,2d-e and 3c). 
Interestingly, the timescales for the emergence of roton peaks in the NS are few ms longer than those observed in the experiment. The origin of this time shift remains an open question, whose answer could e.g. require a refinement of current models of quantum fluctuations. However, despite this delay, the growth rate of the roton population in the NS matches both the experimental observations and the SSM predictions in the early dynamics (Fig.\,3b).

\begin{figure}
\includegraphics{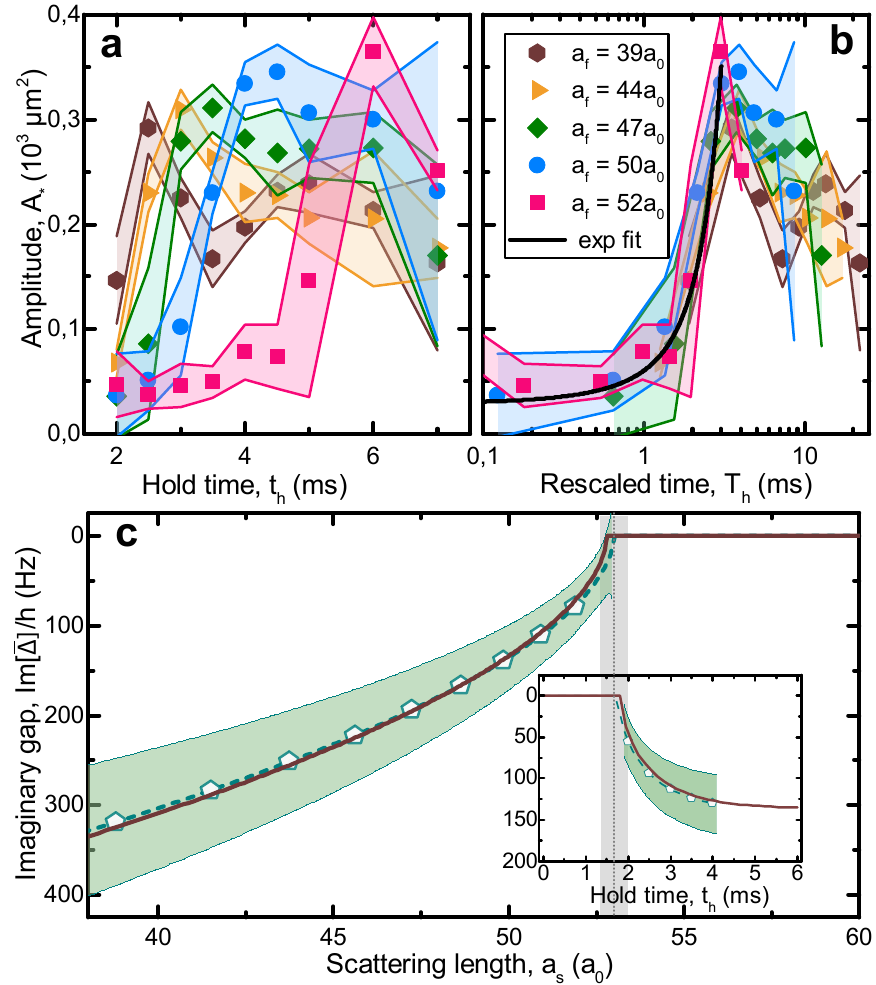} 
\caption{\label{fig4} 
{\bf Population growth and roton gap}. Results for $(\nu_z,\lambda)=(149\,{\rm Hz},4.3)$. {\bf a, b}, $A_*$ depending on $\tho$ and on the rescaled time $T_{\rm h}$, after quenching to $\af=39\,a_0$ (hexagons), $\af=44\,a_0$ (triangles), $\af=47\,a_0$ (diamonds), $\af=50\,a_0$ (circles) and $\af=52\,a_0$ (squares). Shaded areas show {the 95\% confidence interval from the three-Gauss fit}. The black  line shows an exponential fit to the full dataset with $T_{\rm h}<3\,$ms. {\bf c}, Extracted $\Im[\bar\Delta]$ depending on $\as$ and on $\tho$ (inset) from the experiments (dashed line with pentagons) with the propagated errors {from the time-rescaling analysis and exponential fit (see b)} (shaded area), and from the SSM (solid line). The dotted line and the corresponding shaded area show the experimental $\ar$ and its fit's confidence interval. The inset shows the same configuration as Fig.\,3 ($\af=50a_0$).}
\end{figure}

In the experiment, we study the time evolution of the roton mode in a fixed geometry $(\nu_z,\lambda)$. In a first set of measurements, we fix $\af$ and follow the dynamics by recording the momentum distribution at various $\tho$. 
We observe that $\krot$ does not change significantly while the roton population initially grows, in excellent agreement with the theories  (Fig.\,3b-c). The growth rate is a particularly relevant quantity as it is directly connected to the imaginary excitation energy in the Bogoliubov description (for the roton, its gap), as revealed by our theory.
In a second set of experiments, we thus systematically study the growth rate of the roton population for various $\af$ (Fig.\,4). 
Our data show that the roton mode begins to populate from a critical value of the scattering length, $\ar$. 
For $\af> 52a_0$, we do not observe the roton peaks at any time.
For $\af\lesssim 52\,a_0$, after a time delay,  $A_*$ undergoes an abrupt increase. 
At longer time, $A_*$ then saturates and eventually slowly decreases while atoms are coincidentally lost (Fig.\,4a). By further lowering $\af$, the roton population exhibits a faster growth rate and shorter time delay.

From the growth rate of the roton population we now extract an overall roton gap $\bar\Delta(\as)$~(Methods). 
In brief, the dynamical variation of the gap is determined by the leading time dependence of $\as$ and $A_*(\af,\tho)\propto \exp\left(2\int_0^{\tho} dt\Im[\bar\Delta(\as(t))]/\hbar \right)$. 
To investigate the scaling $\bar\Delta(\as)$, we first consider the analytical expression of the elementary local roton gap, $\Delta$ as a function of $\as$ (see Eq.\eqref{eq:homogeneous_spectrum}). Developing $\Delta$ in the vicinity of $\as = \ar$  yields $\Delta(\as) \propto \left(\ar-\as\right)^{1/2}$. Here we confirm this scaling by considering the generic power-law dependence 
$\Im[\bar\Delta(\as)]/\hbar= \Gamma\delta(\as)$ with $\delta(\as)=\left(\frac{\ar-\as}{a_0}\right)^\beta \left(\as\leq \ar\right)$. 
Consequently, $A_*$ grows as $\exp\left(2\Gamma\int_0^{\tho} \delta(\as(t))\,dt\right)\,=\,\exp\left(2\Gamma T_{\rm h}\right)$. By rescaling the time variable as $\tho \rightarrow T_{\rm h}=\int_0^{\tho} \delta(\as(t)) dt$, we determine the parameters $\ar$ and $\beta$ such that all the curves of Fig.\,4a fall on top of each other (Fig.\,4b).  
The best overlap of the experimental curves is found 
for $\ar=53.0(4)\,a_0$ and $\beta=0.55(8)$. We then perform an exponential fit to all the rescaled data for $T_{\rm h}<3\,$ms and extract $\Gamma=465(83)\,{\rm s}^{-1}$.  
The same analysis, applied to the calculated population growth from the SSM for different $\af$, gives $\ar\simeq 52.8\,a_0$, $\beta\simeq 0.55$, and $\Gamma \simeq 472\,{\rm s}^{-1}$, which are very close to the experimental values. This time-resolved study allows to readily extract the imaginary roton gap, $\Im[\bar\Delta]$, from both the experiment and the SSM, as a function of $\as$ and $\tho$ (Fig.\,4c). Our observations show the softening of the roton mode at $\as=\ar$ and the expected increase of $\Im[\bar\Delta]$ for $\as<\ar$, nicely grasping the dynamics during the growth of the roton population (inset of Fig.\,4c). The observed behaviour shows again a remarkable agreement with the theory predictions.

To conclude, our work demonstrates the power of weakly-interacting dipolar quantum gases to access the regime of large-momentum, yet low-energy, excitations dressed by interactions. This newly accessible regime, which is largely unexplored in ultracold gases, raises fundamental questions and opens novel directions. Future key developments are to study the impact of such low-lying excitations on the superfluid behavior of a dBEC~\cite{Lasinio:2013,Blakie:2012,Wilson:2010,Pitaevskii:2016}, the interplay between phononic and rotonic modes in the stability diagram of the quantum gas~\cite{Ronen:2007,Martin:2012}, and the possible role of roton excitations as  triggering mechanism of the recently observed instability leading to the formation of metastable droplet arrays in elongated traps~\cite{Igor:2016,Wenzel:2017}. Of particular interest is the prospect of creating a supersolid and striped ground-states in dBECs~\cite{Boninsegni:2012,Wenzel:2017}. Indeed, the short-wavelength density modulation tied to the roton softening together with the quantum stabilization mechanism can favor the formation of such an exotic phase of matter, in which crystalline order coexist with phase coherence.  Contrary to recent experiments~\cite{Li:2017,Leonard:2017}, where the density modulation is imposed by external fields yielding a supersolid-like arrangement of infinite stiffness, a dipolar supersolid would be compressible. 
Experiments on dBECs provide hence the exciting opportunity to unveil similarities and differences among complementary approaches to supersolidity.



\bibliography{Roton}

\begin{itemize}
 \item We are particularly grateful to B. Blakie for many inspiring exchanges. We thank D. O'Dell, M. Baranov, E. Demler, A. Sykes, T. Pfau, I. Ferrier-Barbut and H. P. B\"uchler for fruitful discussions, and G. Natale for his support in the final stage of the experiment.  This work is dedicated in memory of D.\,Jin and her inspiring example.  
The Innsbruck group is supported through 
a ERC Consolidator Grant (RARE, no.\,681432) and a FET Proactive project (RySQ, no.\,640378) of the EU H2020. LC is supported within a Marie Curie Project (DipPhase, no.\,706809) of the EU H2020. FW and LS thank the DFG (SFB 1227 DQ-mat). All the authors thank the DFG/FWF (FOR 2247). Part of the computational results presented have been achieved using the HPC infrastructure LEO of the University of Innsbruck. 
 \item {\bf Competing Interests:} 
 The authors declare that they have no
competing financial interests.
 \item {\bf Author Information:} 
 Correspondence and requests for materials
should be addressed to F.F.~(email: francesca.ferlaino@uibk.ac.at).
\item {\bf Author contributions:} FF, LC, DP, GF, MJM, JHB, SB conceived and supervised the experiment and took the experimental data. LC analysed them.  RMWvB developed the BdG calculations. FW, RMWvB and LS performed the real time simulations. LS derived the analytical and SSM. LC, FF, RMWvB and LS wrote the paper with the contributions of all the authors.

\end{itemize}

\section*{Methods}

\subsection{Trapping geometries.}
\label{sec:trap}
{The BEC is confined in a harmonic trapping potential $V(\rvec) = 2m \pi^2 (\nu_x^2 x^2 + \nu_y^2 y^2+ \nu_z^2 z^2)$,  
characterized by the frequencies $(\nu_x,\nu_y,\nu_z)$. The ODT results from the crossing of two red-detuned laser beams of $1064\,$nm wavelength at their respective focii. One beam, called vODTb, propagates nearly collinear to the $z$-axis and the other, denoted hODTb propagates along the $y$-axis. By adjusting independently the parameters of the vODTb and hODTb, we can widely and dynamically control the geometry of the trap (see Supplementary Information). In particular, $\nu_y$ is essentially set by the vODTb power while $\nu_z$ (and $\nu_x$) independently by that of the hODTb. This yields an easy tuning of the  trap aspect ratio, $\lambda=\nu_z/\nu_y$, relevant for our cigar-like geometry.}

{After reaching condensation (see Supplementary Information), we modify the beam parameters to shape the trap into an axially elongated configuration, favourable for observing the roton physics ($\nu_y \ll \nu_x, \nu_z$). The trapping geometries probed in the experiments, whose $(\lambda,\nu_x,\nu_y,\nu_z)$ are reported in Supplementary Table 1,  are achieved by changing the hODTb power with the vODTb power set to 7W so that $\nu_y$ and $\nu_z/\nu_x$ are kept roughly constant. 
Only the green triangle in Fig.\,2d is obtained in a distinct configuration, with a vODTb power of 2W, leading to $(\nu_x,\nu_y,\nu_z)=(156,17,198)\,$Hz and $\lambda=11.6$. The $(\lambda,\nu_x,\nu_y,\nu_z)$ are experimentally calibrated via exciting and probing the center-of-mass oscillation of thermal samples. 
We note that the final atom number $N$, BEC fraction $f$, and temperature $T$ after the shaping procedure depend on the final configurations, as detailed in Supplementary Table 1.}

\subsection{Quench of the scattering length $\as$.}
\label{sec:quench}

To control $\as$ we use a magnetic Feshbach resonance  between $\Er$ atoms in their absolute ground state, which is centered around $B=0\,$G~\cite{Chin2010fri}. The $B$-to-$\as$ conversion has been previously precisely measured via lattice spectroscopy, as reported in Ref.\,\cite{Chomaz:2016}. Errors on $\as$, taking into account statistical uncertainties of the conversion and effects of magnetic field fluctuations (e.g. from stray fields), are of 3-to-5\,$a_0$ for the relevant $\as$ range $27$-$67$\,$a_0$ in this work. 
After the BEC preparation and in order to investigate the roton physics via an interaction quench, we suddenly change the magnetic field set value, $B_{\rm set}$, twice. First we perform the quench itself and abruptly change $B_{\rm set}$ from  $0.4\,$G ($\asi=61\,a_0$) to the desired lower value (corresponding to $\af$) at the beginning of the hold in trap ($\tho=0\,$ms). Second we prepare the ToF expansion and imaging conditions (see Method~\textbf{Imaging procedure}) and abruptly change $B_{\rm set}$ from the quenched value back to $0.3\,$G ($\as=57\,a_0$) at the beginning of the ToF expansion ($\ttof=0\,$ms).
 Due to delays in the experimental setup, e.g. coming from eddy currents in our main chamber, the actual $B$ value felt by the atoms responds to a change of $B_{\rm set}$ via $B(t)=B_{\rm set}(t)+\tau dB/dt$ \cite{Lahaye:2008}. By performing pulsed-radio-frequency spectroscopy measurements (pulse duration 100\,$\mu$s) on a BEC after changing $B_{\rm set}$ (from 0.4 to 0.2\,G), we verify this law and extract a time constant $\tau =0.98(5)\,$ms.
 Then $\as$ is also evolving during $\tho$ and $\ttof$ on a similar timescale. This effect is fully accounted for in the experiments and simulations, and we report the roton properties as a function of the effective value of $\as$ at  $\tho$.
We use $\tho$ ranging from $2$ to $7$\,ms. The lower bound on $\tho$ comes from the time needed for $\as$ to effectively reach the regime of interest. We then consider the initial evolution for which $\tho/\nu_y\, ;\,\tho/\tau_{\rm coll} \ll 1$,  
$1/\tau_{\rm coll}$ being the characteristic collision rate. 
We estimate that $\tau_{\rm coll}$ ranges typically from 40 to 90\,ms in the initial BECs of Supplementary Table 1 at $\asi=61\,a_0$. 
Experimentally we observe that the roton, if it ever develops, has developed within the considered range of $\tho$. 

\subsection{Imaging procedure.}
\label{sec:imaging}

In our experiments, we employ ToF expansion  measurements, accessing the momentum distribution of the gas \cite{Pitaevskii:2016}, to probe the roton mode population. We let the gas expand freely for $\ttof=30\,$ms, which translates the spatial imaging resolution ($\sim\,3.7\,\um$) into a momentum resolution of $\sim0.32\,\um^{-1}$ while our typical roton momentum is $\kr \gtrsim 2\,\um^{-1}$. 
After ToF expansion, we record 2D absorption pictures of the cloud  by means of standard resonant absorption imaging on the atomic transition at 401\,nm. The imaging beam propagates nearly vertically, with a remaining angle of $\sim 15^{\rm o}$ compared to the $z$-axis within the $xz$-plane. Thus the ToF images essentially probe the spatial density distribution $\ntof \left(x,y,\ttof\right)$ in the $xy$ plane. When releasing the cloud ($\ttof=0\,$ms), we change $B$ back 
to $B=0.3\,$G (Method \textbf{Quench of the scattering length $\as$}). This change  
enables constant and optimal imaging conditions with a fixed probing procedure, i.\,e.\,, a maximal absorption cross-section. 
In addition, the associated increase of $\as$ to $57\,a_0$ allows to minimize the time during which the evolution happens in the small-$\as$ regime where roton physics develops, such that we effectively probe only the short-time evolution of the gas. 
In this work, we use the simple mapping:
\begin{equation}
\label{eq:ntof}
    \tilde{n}(k_x,k_y) = \left(\frac{\hbar \ttof}{m}\right)^2 \ntof \left(\frac{\hbar k_x\ttof}{m},\frac{\hbar k_y \ttof}{m},\ttof\right),
\end{equation}
which neglects the initial size of the cloud in trap and the effect of interparticle interactions during the free expansion.
Using real-time simulations (see Supplementary Information), we simulate the experimental sequence and are able to compute both the real momentum distribution from the in-trap wavefunction and the spatial ToF distribution $30\,$ms after switching off the trap. Using the mapping of Eq.\,\eqref{eq:ntof} and our experimental parameters, the two calculated distributions are very similar, and, in particular, the two extracted momenta associated with the roton signal agree within $5\%$. This confirms that the interparticle interactions play little role during the expansion and justifies the use of Eq.\,\eqref{eq:ntof}.

\subsection{Fit procedure for the ToF images.}
\label{sec:fitToF}
For each data point of Figs.\,2-4, we record between 12 and 25 ToF images. By fitting a two-dimensional Gaussian distribution to the individual images, we extract their origin $(k_x,k_y)=(0,0)$ and recenter each image. From the recentered images, we compute the averaged $\tilde{n}(k_x,k_y)$, from which we characterise the linear roton developing along $k_y$. To do so, we extract a one-dimensional profile $\tilde{n}_1(k_y)$ by averaging $\tilde{n}(k_x,k_y)$ over $k_x$ within $|k_x|\leq k_{\rm m}= 3.5\um^{-1}$ : 
$\tilde{n}_1(k_y)=\int_{-k_{\rm m}}^{k_{\rm m}} \tilde{n}(k_x,k_y)dk_x/\int_{-k_{\rm m}}^{k_{\rm m}}dk_x$.
To quantitatively analyse the observed roton peaks, we fit a sum of three Gaussian distributions to $\tilde{n}_1(k_y)$.
One Gaussian accounts for the central peak and its centre is constrained to $k_0\sim 0$. Its amplitude is denoted $A_0$. The two other Gaussians are symmetrically located at $k_0\pm k_y^*$, and we constrain their sizes and amplitudes to be identical, respectively equal to $\sigma^*$ and $A_*$ . We define the roton population contrast via $\crot = A_*/A_0$.
We focus on the roton side peaks by constraining $k_y^*>0.5\,\um^{-1}$ and $\sigma^*<3\,\um^{-1}$ (peak at finite momentum of moderate extension compared to the overall distribution).

In order to analyse the onset and evolution of the 
roton population (see Fig.4), we perform a second run of the fitting procedure, in which we constrain the value of $k_y^*$ more strictly. 
The interval of allowed values is defined for each trapping geometry based on the results of the first run of the fitting procedure. We use the results of the ($\af$,\,$\tho$)-configurations where the peaks are clearly visible and we set the allowed $k_y^*$-range to that covered by the 95\% confidence intervals of the first-fitted $k_y^*$ in these configurations. This constraint enables that the fitting procedure estimates the residual background population on the relevant momentum range for the roton physics even for $\as>\ar$ (see e.g. Fig.4a).

\subsection{Time-rescaling analysis and roton gap estimate.}
\label{sec:rescaling} 

 In Fig.\,4a-b, we systematically analyse the time evolution of the roton population for various $\af$ and link it to the roton spectrum in a quench picture. The roton population is embodied by the amplitude $A_*$ of the three-Gaussian fit (Method \textbf{Fit procedure for the ToF images}), which measures the density $\tilde{n}_{1}(\kr)$. $A_*$ is observed to initially  increase if $\af<\ar$, its growth rate increases for decreasing $\af$ and $\ar$ depends on the trap geometry.
 
 If dynamically unstable, the $\krot$ component is expected to grow exponentially with an instantaneous rate $2\Im[\epsilon(\krot,\tho)]/\hbar$. We hence expect an initial growth of $A_*$ of the form: 
\begin{equation}
 \label{eq:A_Delta}
 A_*(\af,\tho)\propto\exp\left(2\int_0^{\tho}dt\Im[\bar\Delta(\af,t)]/\hbar \right),
\end{equation} 
$\bar\Delta(\af,\tho)$ being the instantaneous value of the overall roton gap, corresponding to an average over the cloud, after quenching $\as$ to $\af$. 
Our results show that the most relevant effect of the quench on the roton spectrum is given by the reduction of $\as$ itself. Hence the time dependence of $\bar\Delta$ is determined by $\as(t)$ and, by monitoring $ A_*(\af,\tho)$, one can readily extract the scattering-length dependent gap $\bar\Delta(\as)$, $\bar\Delta(\af,t)=\bar\Delta(\as(t))$. To investigate the scaling $\bar\Delta(\as)$, we first consider the simpler case of the static and local roton gap $\Delta$. From Eq.\,\eqref{eq:homogeneous_spectrum} of the main text characterising this elementary gap, assuming $n_0$, $X$ and $Z$ fixed, and using the fact that  $\Delta(\as=\ar)=0$, one easily 
obtain that for $\as$ in the vicinity of $\ar$, $\Delta(\as) \propto (\ar-\as)^{1/2}$. Here we verify this scaling for the global and dynamical quantity $\bar\Delta(\as)$, considering the generic power-law dependence $\Im[\Delta(\as)]/\hbar= \Gamma\delta(\as)$ with $\delta(\as)=\left(\frac{\as-\ar}{a_0}\right)^\beta \left(\as(t)\leq \ar\right)$, in which the parameters $\Gamma$, $\ar$ and $\beta$ mainly depend on the trap geometry.

 Our full set of data $A_*(\af,\tho)$ for the geometry $(\nu_z,\lambda)=(149\,{\rm Hz},4.3)$ enables us to assess this scaling. Indeed Eq.\,\eqref{eq:A_Delta} then reads:
 \begin{equation}
 \label{eq:A_delta}
 A_*(\af,\tho)\propto\exp\left(2\Gamma\int_0^{\tho}\delta(\as(t))dt\right).
\end{equation}
 This defines a time rescaling $\tho \rightarrow T_{\rm h}=
 \int_0^{\tho}\delta(\as(t))dt$ along which all our experimental data $A_*(\as,\tho)$ should collapse in a unique curve, marked by an initial exponential growth of rate $2\Gamma$. The relevant values of $\ar$, $\beta$ and $\Gamma$ are the one that result in the best overlap of the data for the initial growth of $A_*$ (minimal spread in $T_{\rm h}$).  
 
 In order to determine $\ar$ and $\beta$, we then plot our full dataset as a function of $T_{\rm h}$ for various trial values of these parameters and evaluate the relevance of the trial couple $(\ar,\beta)$. 
 Precisely, we assess the dispersion in $T_{\rm h}$ of the full dataset for few fixed $A_*= A_*^{i}$. We use a panel of 10 values $A_*^{i}$ within 
 [80,200]$\um^2$, 
 corresponding to the range of the initial growth  of $A_*$ for the geometry of Fig.\,4a-b.
 We interpolate the experimental data $A_*(T_{\rm h})$ for each $\af$ using piecewise cubic polynomial interpolation, extract the corresponding set of $T_{\rm h}^{i}$ at which $A_*(T_{\rm h}^{i})=A_*^{i}$. For each $i$, we evaluate the spread of the $T_{\rm h}^{i}(\af)$ by two complementary quantities: (a) the square-root of their variance and (b) the discrepancy between the maximal and minimal value of the set. We finally estimate the accuracy of a unified dependence $A_*(T_{\rm h})$ for a given ($\ar,\beta$) via the geometrical average of (a) and (b) for all the ten $A_*^{i}$ values. We fit the relevant $\ar$ and $\beta$ by minimizing this averaged spread. For the experimental data of  Fig.\,4a-b, this results in $\ar=53.0(4)a_0$ and $\beta=0.55(8)$. 
 
 Using these values of $\ar$ and $\beta$, we observe that the initial growth of $A_*$ extends for $T_{\rm h}\leq 3\,$ms (before saturating and decreasing for longer $T_{\rm h}$).  We estimate $\Gamma$ by performing an exponential fit on the full set of data $A_*(T_{\rm h})$ with $T_{\rm h}\leq 3\,$ms. This gives $\Gamma=465(83)\,{\rm s}^{-1}$. Note that Fig.\,4a-b only show 5 values of $\af$ while the reported analysis considers all available experimental data, i.e. 11 values between $31\,a_0$ and $52\,a_0$. 
 
 From the formula with the estimated $\ar$, $\beta$ and $\Gamma$, we can then compute the global roton gap $\bar\Delta$. The extracted value is only meaningful for $\bar\Delta^2<0$. For the experiments, we also restrict its relevance to $T_{\rm h}\leq3\,$ms so that, e.g., in the inset of Fig.\,4c, $\Im[\bar\Delta]$ is shown up to $\tho= 4\,$ms after which the roton population is observed to deviate from the exponential growth picture (see Figs.\,3b and 4a). Note that the $\Im[\bar\Delta]$ estimates at the different $\tho$ are not independent.

 In Fig.\,2, we report on the roton momentum as a function of the system characteristics ($\as$, $\ell_z$). Here we estimate $\kr$ from a $\tho$ value that is individually optimised for each $\as$ and $\ell_z$ investigated (largest visibility). We point out that the selected $\tho$  correspond, in the $A_*$ dynamics, to the late stage of the exponential growth, close to the maximum (i.e. $T_{\rm h}\sim 3\,$ms).  Here, the atom loss remains at a few percents level.  

{In Fig.\,2d, we show the value of $\krot$ at the onset of the population of the roton mode, i.e. at $\as=\ar$. We estimate $\ar$ for each trap geometry by analysing the evolution of the roton population with $\as$. We employ a simplified approach with respect to the one in Fig.\,4 (see Supplementary Information), which we estimate to lead to a maximum underestimate for $\ar$ of about $1.5\,a_0$, lying within our experimental uncertainty on $\as$; see Supplementary Table 2.}

\subsection{Analytical dispersion relation for an infinite axially elongated geometry.}
\label{sec:analyptical}
Equation \eqref{eq:homogeneous_spectrum} in the main text results from a similar procedure as that 
used in Ref.~\cite{Santos:2003} for rotons in infinite q2D traps.
We consider a dBEC homogeneous along $y$ but harmonically confined with frequencies $\nu_x$ and $\nu_z$ along $x$ and $z$.
For sufficiently strong interactions the BEC is in the TF regime on the $xz$ plane, in which the BEC wavefunction acquires the form 
$\psi_0(\brho)=\sqrt {n(\brho)}$, with $n(\brho)=n_0 \left ( 1 - (x/X)^2 - (z/Z)^2 \right )$, where $X$ and 
$Z$ are the TF radii, and $\brho=(x,z)$. The calculation of $n_0$, $X$ and $Z$ is detailed at the end of this section.

Due to the axial homogeneity, the elementary excitations of the Bogoliubov-de Gennes spectrum have 
a defined axial momentum $k_y$, and take the form $\delta\psi(\br,t) = u({\brho})e^{ik_y y-i\epsilon t/\hbar} 
-v({\brho})e^{-ik_y y+i\epsilon t/\hbar}$, {where $u, v$ denote the amplitudes of the spatial modes oscillating in time with characteristic frequency $\epsilon /\hbar$  (see Supplementary Information for more details).}
We consider the standard GPE without LHY correction and three-body 
losses, and insert the perturbed solution
$\psi(\br,t)=\left ( \psi_0(\brho) + \eta \delta\psi(\br,t)  \right 
)e^{-i\mu t/\hbar}$ {, where $\mu$ is the chemical potential associated to $\psi_0$ and $\eta \ll 1$}. After linearization we obtain the BdG equations for 
$f_\pm(\brho)=u(\brho)\pm v(\brho)$:
\begin{eqnarray}
\epsilon f_ -(\brho)&=& H_{\rm kin}f_+(\brho), \\
\epsilon f_ +(\brho)&=& H_{\rm kin}f_-(\brho) + H_{\rm int}[f_-(\brho)],
\end{eqnarray}
where
\begin{eqnarray}
H_{\rm kin} f_\pm(\brho)&=& \frac{\hbar^2}{2m} \left ( 
-\nabla^2+k_y^2+\frac{\nabla^2\psi_0}{\psi_0} \right ) f_\pm(\brho), \\
H_{\rm int} [f_-(\brho)] &=& 2   \int  d^3 r' U(\br-\br') \mathrm{e}^{-\mathrm{i}k_y(y-y')} 
\nonumber \\
&& \,\,\,\,\,\,\,\,\,\,\,  \psi_0(\brho)\psi_0(\brho')f_-(\brho').
\end{eqnarray}
{where $U(\rvec)$ is the binary interaction potential for two particles separated by $\rvec$. $U(\rvec)= g \left(\delta(\rvec) + \frac{3 \epsdd}{4\pi} \frac{1 - 3 \cos^2 \theta}{|\rvec|^3}\right)$ includes  both contact and dipolar interactions (see Supplementary information).}

Employing $f_+(\brho)=W(\brho)\psi_0(\brho)$, and for $k_y\gg1/X,1/Z$, we 
obtain the following equation for the function $W(\brho)$:
\begin{eqnarray}
0&=&2gn_0 \left (1-\tx^2-\tz^2 \right )\left [\frac{1}{X^2} 
\frac{\partial^2W}{\partial\tx^2}+\frac{1}{Z^2}\frac{\partial^2W}{\partial\tz^2} \right ] \nonumber 
\\
&-&gn_0 (1+2\epsilon_{dd})\left [\frac{1}{X^2} \tx \frac{\partial 
W}{\partial\tx}+\frac{1}{Z^2} \tz \frac{\partial W}{\partial\tz} 
\right ] \nonumber \\
&+& \left ( \frac{2m}{\hbar^2} (\epsilon^2-E(k_y)^2)-2g\epsilon_{dd} 
n_0 \left (\frac{1}{X^2}+\frac{1}{Z^2} \right ) \right ) \nonumber \\
&-& \frac{4m}{\hbar^2}gn_0(1-\epsilon_{dd})E(k_y) \left ( 
1-\tx^2-\tz^2 \right ),
\label{eq:W}
  \end{eqnarray}
where $\tx=x/X$, $\tz=z/Z$, $E(k_y)=\hbar^2 k_y^2/2m$. For 
$\epsilon_{dd}=1$ the last term of Eq.~\eqref{eq:W} vanishes. In that case, the 
lowest-energy solution is given by $W=1$, whose eigen-energy builds, as a function of $k_y$, 
the dispersion $\epsilon_0(k_y)$ with 
\begin{equation}
\epsilon_{0}(k_y)^2=E(k_y)^2+ E_0^2,
\end{equation}
with $E_0^2=2 g \epsilon_{dd}n_0  \frac{\hbar^2}{2m} 
\left ( \frac{1}{X^2} + \frac{1}{Z^2} \right )$.
In the vicinity of $\epsdd=1$, the effect of the last term in 
Eq.~\eqref{eq:W} may be evaluated perturbatively, resulting in the dispersion
\begin{equation}
\epsilon(k_y)^2\simeq \epsilon_0(k_y)^2- 2 E_I E(k_y),
\label{eq:W-2}
\end{equation}
with $E_I=\frac{2}{3}gn_0(\epsilon_{dd}-1)$. 

This expression for the dispersion presents a roton minimum for $\epsilon_{dd}>1$ at 
$\kr= \frac{1}{\hbar}\sqrt{2mE_I}$. Expanding Eq.~\eqref{eq:W-2} in the vicinity of the roton minimum, 
$\epsilon(k_y)^2\simeq \epsilon(\kr)^2 + \frac{1}{2}\left [  \frac{d^2\epsilon^2(k_y)}{dk_y^2} \right ]_{k_y=\kr}$, 
we obtain Eq. \eqref{eq:homogeneous_spectrum} of the main text, with $\Delta=\epsilon(\kr)=\sqrt{E_0^2-E_I^2}$. At the instability, $\Delta= 0$, 
and $\kr= \frac{1}{\hbar}\sqrt{2mE_0}$.

Employing a similar procedure  as in Ref.~\cite{Odell2003} we obtain that the BEC aspect ratio
$\chi = Z/X$ fulfills:
\begin{equation}
\chi^2\left [  
\frac{(1-\epsilon_{dd})(1+\chi)^2+3\epsilon_{dd}}{(1+2\epsilon_{dd})(1+\chi)^2-3\epsilon_{dd}\chi^2}\right  
]
= \lp^2,
\end{equation}
with $\lp=\nu_x/\nu_z$ and
\begin{equation}
Z^2=\frac{gn_0}{2\pi^2 m \nu_z^2}\left [  
(1+2\epsilon_{dd})-\frac{3\epsilon_{dd}\chi^2}{(1+\chi)^2}\right ]
\end{equation}
These two equations fully determine the TF solution for given  
$\epsilon_{dd}$, $gn_0$, and $\lp$. 
By inserting the expressions of $X^2$ and $Z^2$ in $E_0$, we find for  $\epsdd\simeq 1$:
\begin{equation}
E_0^2 = \frac{h^2\nu_z^2}{6} (1+\chi)^2 \left(\lp^2+\frac{1}{1+2\chi}\right)
\end{equation}
whereas $\chi$ simplifies into $\chi=\lp(1+\sqrt{1+1/\lp})$. 
As a result, at the instability $\kr \ell_z$ depends only on the transverse confinement aspect ratio $\lp$, giving the geometrical factor $\kappa$:  
\begin{equation}
\kappa =\kr\ell_z = \left(\frac{2}{3}\right)^{1/4} \sqrt{1+\chi} \left(\lp^2+\frac{1}{1+2\chi}\right)^{1/4}.
\end{equation}

\subsection{Self similar model and instantaneous roton spectrum.}
\label{sec:analdyn}

The dynamics of the condensate during and after the quench is crucial to understand the resulting momentum peaks and their direct relation to the growth of roton excitations. Due to its large 
characteristic momentum, the roton spectrum may be evaluated at any time using local density approximation. On the other side, the evolution of the local density is determined by the global dynamics of the condensate induced by the quench. Hence, interestingly, the analysis of the effect of the quench on the roton spectrum may be performed by combining a self-similar theory of the global dynamics~\cite{Giovanazzi:2006}, describing the evolution at low $k_y$, and the model of the roton spectrum developed in the Method \textbf{Analytical dispersion relation for an infinite axially elongated geometry}, accounting for the  high-$k_y$ region of interest.  
We note that this analysis approximates the real time evolution provided by the standard NLGPE~\cite{Baranov:2008,Pitaevskii:2016,Blakie:2008}. 
This treatment is valid for moderate-enough quenches so that each of the descriptions applies~(see below for an in-depth discussion) and as long as the high-$k$ modes population only minimally impact the BEC dynamics, that is for short-enough 
time scales.

We assume that the condensate preserves its TF shape during the evolution:
\begin{equation}
 n({\mathbf r},t)
 =n_0(t)\left [ 1-\left (\frac{x}{X(t)}\right )^2-\left (\frac{y}{Y(t)}\right )^2-\left (\frac{z}{Z(t)}\right )^2 \right ],
\end{equation}
where $X(t)=b_x(t)X_0$, $Y(t)=b_y(t)Y_0$, and $Z(t)=b_z(t)Z_0$ are the re-scaled TF radii, with $b_{x,y,z}(t)$ the scaling coefficients~($b_{x,y,z}(0)=1$). {Their evolution after the quench can be deduce from solving the hydrodynamics equations (see Supplementary information).} 
Prior to the quench of $\as$, the stationary TF solution is obtained from the self-consistent equations $\chi_x(0)=\frac{\omega_x}{\omega_z}\sqrt{C(0)/A(0)}$ and 
$\chi_y(0)=\frac{\omega_y}{\omega_z}\sqrt{C(0)/B(0)}$. Solving these equations and using normalisation provides $X(0)$, $Y(0)$, $Z(0)$, and $n(0)$ for known numbers of atoms $N$ and 
trap frequencies $\omega_{x,y,z}$. We use this stationary solution as the initial condition at the start of the quench~($t=0$), and solve the 
system of differential equations resulting from the hydrodynamics evolution to obtain the scaling coefficients. An example for the relevant parameters of Fig.\,3 is shown in  Supplementary Fig.\,3.

For a given position $y$ we then evaluate the local TF profile:
$n({\mathbf r},t)=n_0(y,t)\allowbreak\big [ 1-(x/X(y,t))^2-(z/Z(y,t))^2 \big ]$, where 
$n_0(y,t)=n_0(t)\left [ 1-(y/Y(t))^2 \right ]$, $X(y,t)=X(t)\left [ 1-(y/Y(t))^2 \right ]^{1/2}$, and $Z(y,t)=Z(t)\left [ 1-(y/Y(t))^2 \right ]^{1/2}$, 
with $n_0(t)=n_0(0)/b_xb_yb_z$. We may then employ a local density approximation and evaluate the local~(in $y$) instantaneous roton spectrum 
using the results of the Method \textbf{Analytical dispersion relation for an infinite axially elongated geometry}:
\begin{eqnarray} 
\frac{\epsilon^2(k_y,y,t)}{(\hbar\omega_z)^2 l_z^4}&\simeq& \frac{k_y^4}{4}-\frac{8\pi}{3}n_0(y,t)(a_{dd}-\as(t))k_y^2 \nonumber \\
&+& 4\pi n_0(y,t)a_{dd}\left [\frac{1}{X(y,t)^2}+\frac{1}{Z(y,t)^2} \right ].
\end{eqnarray}
The roton population grows exponentially when the spectrum becomes imaginary, leading to the appearance of the corresponding momentum 
peaks at large momenta. From the local instantaneous roton spectrum we may estimate the momentum peak as
\begin{equation}
 \tilde{n}_1(k_y,t) \sim \int dy  \, \left [ e^{2\int_0^t dt' \Im\left [\epsilon(k_y,y,t')\right]}-1\right ].
\end{equation}
where we assume equal seeding 
for all unstable modes. We then evaluate the total population of the peak $N_{*}(t)=\int d k_y  \tilde{n}_1(k_y,t)$, as well as the 
roton momentum $\langle k \rangle = \int dk_y k_y \frac{\tilde{n}_1(k_y,t)}{N_{*}(t)}$. 

The previously discussed SSM does not describe properly the evolution when the shrinking of the condensate, 
and hence the increase of its peak density, is too large. As the gas gets dense, quantum fluctuations, which are not considered 
in the SSM, introduce in our experiments an effective repulsion that crucially prevents large densities or 
condensate collapse (see Supplementary Information). This limits the use of the SSM to the vicinity of $\ar$. Moreover, for our tightest trap 
the SSM results are also unreliable, since the theory predicts condensate collapse before the momentum peak develops.
That said, as shown in the main text, the SSM provides 
not only a clear intuitive understanding of our measurements, but to a very large extent also a very good quantitative 
agreement with both our experimental results and our NS based on a generalized GPE, 
which is detailed in Supplementary Information.


{\bf Data availability:} The data that support the plots within this paper and other findings of this study are available from the corresponding author upon reasonable request.

\newpage

\section*{Supplementary Information}

\subsection{Production of $\Er$ BECs.}
\label{sec:bec}
We prepare a $\Er$ BEC similarly to Refs.\,\cite{Chomaz:2016,Aikawa:2012}. From a narrow-line magneto-optical trap with $3\times10^7$ \Er\ atoms, automatically spin-polarized in their absolute lowest Zeeman sub-level \,\cite{Frisch:2012}, at about $10\mu{\rm K}$, we directly load the atomic gas in a crossed optical dipole trap (ODT) with an efficiency of more than $30\%$.
A uniform  magnetic field, $B$, is permanently applied along the vertical $z$ axis, fixing the dipole orientation, while its value is varied during the experimental sequence, to tune $\as$ (see Method \textbf{Quench of the scattering length $\as$}). We achieve condensation by means of  evaporative cooling  in the crossed ODT at $B =1.9\,$G ($\as =80(2)\,a_0$). During the evaporation procedure, we first change the power and then the ellipticity of one of the ODT beams (see Section \ref{sec:trap2}). The final atom number, typically $10^5$, and condensed fraction, typically $70\%$,  
are assessed by fitting the time-of-flight (ToF) absorption images of the gas to a bimodal distribution, sum of a TF profile and a broad Gaussian background.

\subsection{Details on the trapping geometries.}
\label{sec:trap2}
The harmonic potential $V(\rvec) = 2m \pi^2 (\nu_x^2 x^2 + \nu_y^2 y^2+ \nu_z^2 z^2)$ trapping the \Er atoms results from the crossing of two red-detuned laser beams of $1064\,$nm wavelength at their respective focii; the hODTb propagating along the $y$-axis and the vODTb, propagating nearly collinear to the $z$-axis. The vODTb has a maximum power of 7\,W and an elliptical profile with waists of $110$ and $55\,\mu{\rm m}$ along $x$ and $y$ respectively. The hODTb has a maximum power of 24\,W, a vertical waist $w_z=18\,\um$, and a controllable horizontal waist, $w_x= \lambda' w_z$. The ellipticity $\lambda'$ can be tuned from $1.57$ to $15$ by time averaging the frequency of the first-order deflection of an Acousto-Optic Modulator \cite{Ahmadi:2005}. By adjusting independently $\lambda'$ and the powers of the vODTb and of the hODTb, we can control the geometry of the trap. Precisely, $\nu_y$ is essentially set by the vODTb power, $\nu_z$ by that of the hODTb, and $\nu_x$ is controlled by both the power and ellipticity of the hODTb, with $\nu_z/\nu_x \approx \lambda'$. 

We use the tuning of the hODTb power and ellipticity to perform evaporative cooling to quantum degeneracy (see Section \ref{sec:bec}). After reaching condensation, we again modify the beam parameters 
to shape the trap into an axially elongated configuration ($\nu_y \ll \nu_x, \nu_z$). The trapping geometries probed in the experiments,  $(\lambda,\nu_x,\nu_y,\nu_z)$, are reported in Supplementary Table 1. They are achieved by changing the hODTb power with $\lambda'=1.57$ and the  vODTb power set to its maximum so that $\nu_y$ and $\nu_z/\nu_x$ are kept roughly constant. 
Only the green triangle in Fig.\,2d is obtained in a distinct configuration, with a vODTb power of 2W, leading to $(\nu_x,\nu_y,\nu_z)=(156,17,198)\,$Hz and $\lambda=11.6$. The $(\lambda,\nu_x,\nu_y,\nu_z)$ are experimentally calibrated via exciting and probing the center-of-mass oscillation of thermal samples. 
We note that the final atom number $N$, BEC fraction $f$, and temperature $T$ after the shaping procedure depend on the final configurations, as detailed in Supplementary Table 1. $T$ is extracted from the evolution with the ToF duration $\ttof$ of the size of the background Gaussian in the TF-plus-Gaussian bimodal fit to the corresponding ToF images of the gas. The values of $N_{\rm c}=fN$ and $T$ are used for  
the initial states $\psi_i$ of our real-time simulations (see Section \ref{sec:realtime}).

\begin{table}[ht!]
\centering
\caption{dBEC parameters 
for the experimental measurements (Figs.\,2-4). The typical statistical uncertainties on $\nu_x$ and $\nu_z$ are below $1\,\%$, and can be up to $10\,\%$ for $\nu_y$. The experimental repeatability results in $5$-to-$ 10\,\%$ shot-to-shot fluctuations of $N$, $f$ and $T$.\label{table}}
\medskip
\begin{tabular}{ccccccc}
\hline
  $\lambda$ & $\nu_x$\,(${\rm Hz}$ ) & $\nu_y$\,(${\rm Hz}$ ) & $\nu_z$\,(${\rm Hz}$ ) & $N$\,($10^4$) & $f$\,(\%) & $T$\,(nK)\\
\hline
	$4.3$	&  $114$ &	$35$ &	$149$ &	$9$ &	$66$ & $45$\\
	$4.9$	&  $154$ &	$40$ &	$195$ &	$10$ &	$65$ & $50$\\
	$10.2$ &	$183$ &	$30$ &	$306$ &	$11$ &	$62$ &	$104$\\
	$14.3$ &	$267$ &	$32$ &	$456$ &	$8.6$ &	$50$ &	$150$\\
	$21.3$ &	$357$ &	$30$ &	$638$ &	$8.4$ &	$36$ &	$179$\\
	$29.7$ &	$432$ &	$26$ &	$771$ &	$7$ &	$20$ &	$171$\\

\hline
\end{tabular}
\end{table}

\subsection{$\boldsymbol \ar$ for various trap geometries.}
\label{sec:astar}

We estimate $\ar$ for each trap geometry of Fig.\,2d by studying the roton population as a function of $\as$.
When decreasing $\as$, the roton population and in particular its contrast $\crot= A_*/A_0$ (see Method \textbf{Fit procedure for the ToF images}) exhibits a sharp increase from an essentially zero and flat contrast. We extract $\as^{*,c}$ as the value of the scattering length corresponding to the onset of the increase of the contrast for $\tho=3\,$ms; see Supplementary Fig.\,1 and Supplementary Table 2. This is a simplified approach with respect to the one in Fig.\,4 (see Method \textbf{Time-rescaling analysis and roton gap estimate.}), which we estimate to lead to a maximum underestimate for $\ar$ of about $1.5\,a_0$, which lies within our experimental uncertainty on $\as$.

\begin{table}[ht!]
\centering
\caption{Roton softening threshold for the geometries realized in the experiment (Table \ref{table}), deduced from the $\crot$ fits for the experimental onset of the population ($\as^{*,\rm c}$) and from imaginary time NS ($\as^{*,\rm st}$). 
The fit uncertainties on the experimental values are typically of 0.2$a_0$, except for the last trap it is of 0.8\,$a_0$. 
\label{table2}
}
\medskip
\begin{tabular}{cccc}
\hline
  $\lambda$ & $\nu_z$\,(${\rm Hz}$ ) & $\as^{*,\rm c}$\,($a_0$) & $\as^{*,\rm st}$\,($a_0$)\\
\hline
	$4.3$   &  $149$ & $51.4$    &  $50$   \\
	$10.2$ &	$306$ &	$50.2$    &   $46$ \\
	$14.3$ &	$456$ &	$49$    &    $43$\\
	$21.3$ &	$638$ &	 $47.7$    &    $39$\\
	$29.7$ &	$771$ &	$45.2$     &     $30$\\
\hline
\end{tabular}
\end{table}

\begin{figure}[ht!]
\includegraphics[width=0.49\textwidth]{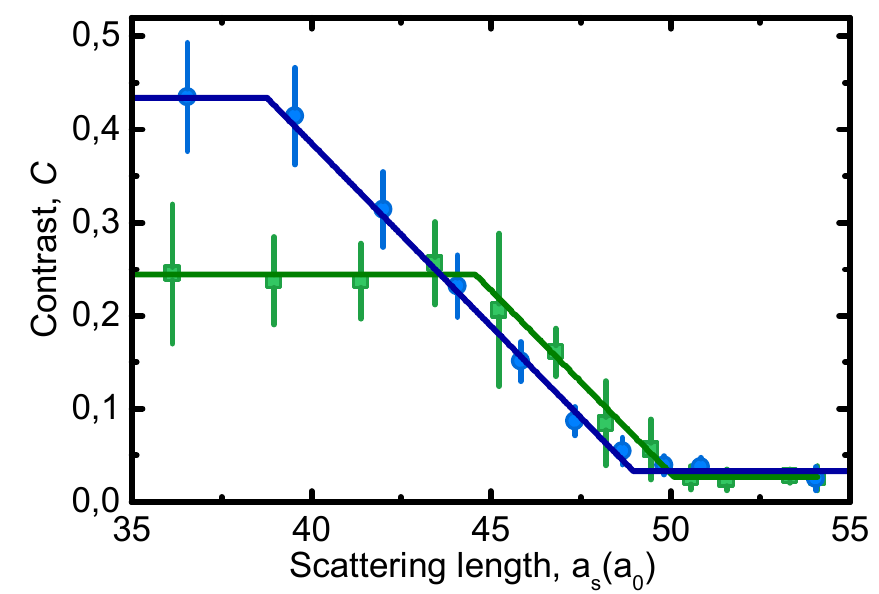} 
\caption{\label{fig4sup}
{\bf Onset of the roton population}. Evolution of $\crot$ with $\as$ at $\tho=3\,$ms, in the geometries $(\nu_z,\lambda)=(149\,{\rm Hz},4.3)$ (green squares) and $(\nu_z,\lambda)=(456\,{\rm Hz},14.4)$ (blue circles). The error bars correspond to the propagated errors from the 95\% confidence interval on $A_{*,0}$ from the three-Gauss fit. We empirically fit a linear step function to identify $\as^{*,c}$ (line). 
} 
\end{figure}

\subsection{Generalized Non-local Gross-Pitaevskii equation.}
\label{sec:nlgpe}
Our theory is based on an extended version of the non-linear Gross-Pitaevskii Equation (NLGPE)
\begin{align} \label{eq:GPE}
{\mathrm i}\hbar\frac{\partial\psi(\rvec, t)}{\partial t}&=\Big(-\frac{\hbar^2 \nabla^2}{2m}+V(\rvec) + \int d\rvec' U({\mathbf r}-{\mathbf r}')n({\mathbf r}') \nonumber\\ & \hspace{2cm}+ \Delta \mu[n] - {\mathrm i}\hbar \frac{L_3}{2}n^2\Big)\psi(\rvec, t) \\
& \equiv \Big(\hat{H}_{\mathrm{GP}}[\psi] - \mathrm{i} \hbar \frac{L_3}{2} n^2\Big) \psi(\rvec, t),\label{eq:GPEcompact}
\end{align}
governing the evolution of a macroscopically occupied  
wavefunction $\psi(\rvec, t)$, with corresponding atomic density $n(\rvec, t) = |\psi(\rvec, t)|^2$ at position $\rvec$ and time $t$. The standard dipolar NLGPE includes the kinetic energy, external trap potential and the mean-field effect of the interactions \cite{Pitaevskii:2016,Baranov:2008}. These correspond to the three first terms of Eq.\,\eqref{eq:GPE}, where the mean-field interaction potential takes the form of a convolution of $n$ with the binary interaction potential
\begin{equation}
U(\rvec) = g \left(\delta(\rvec) + \frac{3 \epsdd}{4\pi} \frac{1 - 3 \cos^2 \theta}{|\rvec|^3}\right),
\end{equation}
for two particles separated by $\rvec$~\cite{Baranov:2008}. The first term corresponds to contact interactions between the particles with strength $g = \frac{4\pi \hbar^2 \as}{m}$. The DDI gives rise to the second term, which depends on both the distance and orientation (angle $\theta$) of the vector $\rvec$ compared to the polarisation axis ($z$ axis) of the dipoles. Most properties of dBECs are well captured by this standard NLGPE (mean-field) \cite{Pitaevskii:2016,Baranov:2008}. 

Recent experimental and theoretical results, however, have established the importance of accounting for quantum fluctuations in dBECs~\cite{Igor:2016, Chomaz:2016, Waechtler:2016,Waechtler:2016b,Bisset:2016}. Their effect can be included in the NLGPE in a mean field treatment through a Lee-Huang-Yang correction to the chemical potential, $\Delta \mu[n] = 32 g (n a_s)^{3/2}(1 + 3 \epsdd^2 / 2) / 3 \sqrt{\pi}$, which is obtained under a local density approximation~\cite{Pelster:2011,Pelster:2012}. The accuracy of this mean field treatment has been established, e.g., in Refs.~\cite{Waechtler:2016,Waechtler:2016b,Bisset:2016}, and has proven succesful in explaining recent experimental results~\cite{Igor:2016, Chomaz:2016}. 
The final nonlinear term in the extended NLGPE accounts for three-body losses~\cite{Kagan:1998}, with an experimentally determined loss parameter $L_3$, which is dependent on $\as$ and typically of the order $L_3 \simeq 10^{-41} \mathrm{m}^6 \mathrm{s}^{-1}$, as reported in Ref.\,\cite{Chomaz:2016}.

\subsection{Bogoliubov-de Gennes spectrum.}
\label{sec:bdg}

Collective excitations of the dBEC are obtained by linearising the NLGPE (see Section \ref{sec:nlgpe}) around a stationary state $\psi_0$, which can be obtained by imaginary time propagation (see Section \ref{sec:realtime}). We write $\psi =  \mathrm{e}^{-\mathrm{i} \mu t/\hbar}(\psi_0 + \eta [u \mathrm{e}^{-\mathrm{i} \epsilon t/\hbar} - v^* \mathrm{e}^{+\mathrm{i} \epsilon t/\hbar}])$, where $\mu$ is the chemical potential associated with state $\psi_0$, and $u, v$ are spatial modes oscillating in time with characteristic frequency $\epsilon /\hbar$  and $\eta \ll 1$  
\cite{Ronen:2006}. Inserting this ansatz in the NLGPE, and retaining only terms up to linear order in $\eta$ we obtain the Bogoliubov-de Gennes (BdG) equations
\begin{equation}
\left(\begin{array}{cc} \hat{H}_{\mathrm{GP}}[\psi_0] + A & -A \\ A & -\hat{H}_{\mathrm{GP}}[\psi_0] - A \end{array}  \right)  \left(\begin{array}{c} u \\ v \end{array}\right)= \epsilon \left(\begin{array}{c} u \\ v \end{array}\right),
\end{equation}    
where the operator $A$, acting on a function $f$ and evaluated at point $\rvec$, is defined as
\begin{multline}
(A f)(\rvec) = \int \mathrm{d}\rvec' \psi_0(\rvec') U(\rvec - \rvec') f(\rvec') \psi_0(\rvec) \\+ \frac{16}{\sqrt{\pi}}g \as^{3/2} \left( 1 + \frac{3}{2} \epsdd^2 \right)|\psi_0(\rvec)|^3 f(\rvec).
\end{multline}
The above equations constitute an eigenvalue problem, which we solve numerically using the Arnoldi method to obtain eigenmodes $(u, v)$ and corresponding excitation energies $\epsilon$. The equations presented here are a generalization of the BdG equations for dipolar systems as derived in Ref. \cite{Ronen:2006}, to include the LHY correction accounting for quantum fluctuations. 

In order to depict the spectrum as a quasi-dispersion relation even in the presence of an axial confinement, we associate to each elementary excitation an effective  momentum $k_y^{(\rm eff)} = \langle k_y^2  \rangle^{1/2}$~\cite{Bisset:2013}.
The spectrum is discrete with phonon-like collective modes  at low $k_y^{(\rm eff)}$. For higher $k_y^{(\rm eff)}$, the spectrum flattens, but eventually bends upwards again due to the  dominant kinetic energy. Instead of developing a smooth minimum,  roton excitations appear as isolated low-lying modes at intermediate momenta that depart from the overall spectrum~\cite{Bisset:2013}.
These so-called roton fingers are related to confinement of the roton modes in the inhomogeneous BEC of profile $n_0(y)$ ~\cite{Lasinio:2013}, see also discussion in the main text.

\begin{figure*}[ht!]
  \centering
\includegraphics[width=0.98\textwidth]{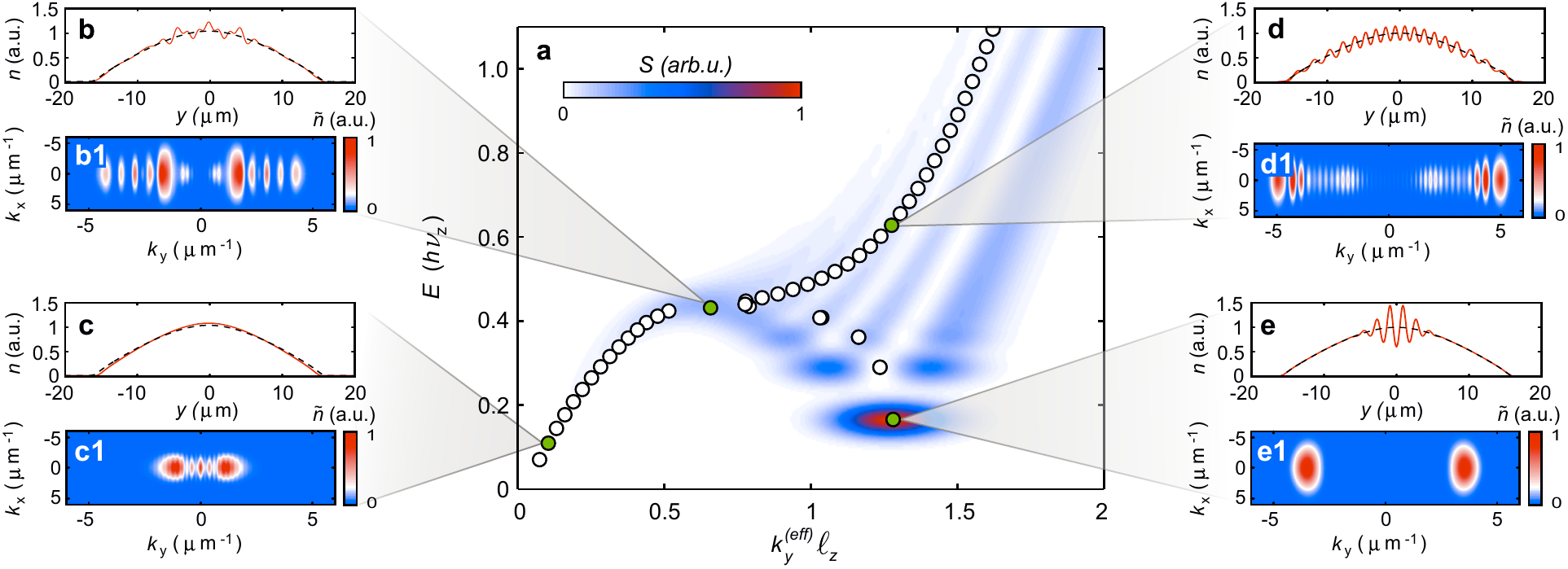} 
\caption{\label{theory}
{\bf BdG excitation spectrum}. {\bf a}, Excitation spectrum of the ground state of a BEC with $N = 50.000$ $\Er$ atoms in a trap with $(\nu_x,\nu_y,\nu_z)=(267,32,456)\,{\rm Hz}$ and scattering length $\as = 43.75\,a_0$, obtained by numerically solving the BdG equations. Roton modes appear as isolated modes lying below the main branch of the spectrum, forming a 'roton finger'. The dynamic structure factor $S$ corresponding to each of the modes is indicated with colored shading. Highlighted in green are several exemplary modes, with panels {\bf b} - {\bf e} showing the corresponding excited state density modulation (black dashed line indicating the ground state, red solid line the excited state), and panels  {\bf b1} - {\bf e1}, the corresponding momentum distribution from $30$\,ms ToF expansion (Methods). {\bf b,~b1} intermediate mode, {\bf c,~c1} phonon mode, {\bf d,~d1} single particle mode, {\bf e,~e1} roton mode.}
\end{figure*}

The confinement is evident from the BdG calculations, in which the lowest roton mode forms a short-wavelength density modulation localized at the trap center (Supplementary Fig.\,\ref{theory}e).
This contrasts with phonon modes for which the modulation is delocalised over the entire condensate (Supplementary Fig.\,\ref{theory}c).
The excited states shown in Supplementary Fig.\,\ref{theory}b-e correspond to the density $|\psi_0 + \eta (u - v^*)|^2$, for particular pairs of $(u,v)$ corresponding to phonon and roton modes (Supplementary Fig.\,\ref{theory}c and e), and exemplary modes at higher energies (Supplementary Fig.\,\ref{theory}b and d). Even while the amplitude $\eta = 0.2$ of the excited modes is taken to be equal in Figs.\,\ref{theory}b - \ref{theory}e, the roton excitation (Supplementary Fig.\,\ref{theory}e) leads to markedly larger local density modulations than the phonon excitation (Supplementary Fig.\,\ref{theory}c). The ToF signatures in Supplementary Fig.\,\ref{theory} are computed by letting the wave function of the excitation, $\eta (u - v^*)$, expand ballistically for $30 \mathrm{ms}$, i.e. neglecting interactions during the expansion. The resulting density $|\eta (u - v^*)|^2$ is then plotted (Supplementary Fig.\,\ref{theory}b1 - e1).

Carrying out the numerical BdG spectrum calculation for various trap parameters, scattering lengths and atom numbers confirms the geometrical scaling $\kr \sim 1 / \ell_z$, as well as its weak dependence on $\as$ close to the instability as expected from the analytical model. We note in particular a good quantitative agreement of $\kr\ell_z$ in Supplementary Fig.\,\ref{theory} with the (stationary) analytic factor $\kappa=1.3$ for this same trap geometry (see Method \textbf{Analytical dispersion relation for an infinite axially elongated geometry}).   A qualitative agreement of the BdG stationary spectrum with the experimental observations is however not fully reached because of the role of the dynamics. A treatment of this effect is provided by the SSM  and NS as discussed in the main text (see also the corresponding Method and Section \ref{sec:realtime}).

Finally, an alternative way of visualizing the excitation spectrum is achieved by computing the dynamical structure factor at $T = 0$ for each mode $(u,v)$ \cite{Zambelli:2000, Blakie:2012},
\begin{equation}\label{EqStrucFac}
S(\bf{k}, \omega') = \sum_j \left|\int \mathrm{d}\rvec [u^*(\rvec) +v^*(\rvec)]\mathrm{e}^{\mathrm{i} \bf{k}\cdot \rvec}\psi_0(\rvec) \right|^2 \delta(\omega' - \omega),
\end{equation}
where $\omega = \epsilon / \hbar$. The structure factor of the axial modes of our finite trapped system (Supplementary Fig.\,\ref{theory}a) presents features resembling a roton-maxon spectrum, its continuum limit being a single-valued spectrum as depicted in Fig.\,1b. To enhance the visibility, the delta function in Eq.\,(\ref{EqStrucFac}) is replaced by a Gaussian with a small finite width in $\omega$. 
Since the dynamical structure factor determines the response of the system when probed at specific energies and momenta, such as e.g. in Bragg spectroscopy experiments~\cite{Brunello:2001, Blakie:2012}, it is interesting to note the difference in amplitude between the roton modes and other parts of the spectrum. In particular, in our quench experiments where the system is effectively driven at a range of energies and momenta, one would expect the strongest response from the roton modes.

\subsection{Self-similar dynamics of the BEC.}

In our self-similar model we account for the dynamics of the BEC profile and its effect on the roton spectrum (see details in the corresponding Method).  To do so, we assume that the condensate preserves its TF shape during the evolution:
\begin{equation}
 n({\mathbf r},t)
 =n_0(t)\left [ 1-\left (\frac{x}{X(t)}\right )^2-\left (\frac{y}{Y(t)}\right )^2-\left (\frac{z}{Z(t)}\right )^2 \right ],
\end{equation}
where $X(t)=b_x(t)X_0$, $Y(t)=b_y(t)Y_0$, and $Z(t)=b_z(t)Z_0$ are the re-scaled TF radii, with $b_{x,y,z}(t)$ the scaling coefficients~($b_{x,y,z}(0)=1$). The corresponding hydrodynamic equations reduce to:
\begin{equation}
\frac{m}{2}\boldsymbol{\nabla} \left [ \frac{\ddot X}{X} x^2+ \frac{\ddot Y}{y} y^2 + \frac{\ddot Z}{Z} z^2 \right ] = -\boldsymbol{\nabla}\mu({\mathbf r},t),
\label{eq:SSM1}
\end{equation}
where $\mu({\mathbf r},t)=V({\mathbf r})+gn({\mathbf r},t)+\int d^3r' V_{dd}({\mathbf r}-{\mathbf r}') n({\mathbf r}',t)$ is the local chemical potential. 
The latter acquires the form:
\begin{equation}
\mu({\mathbf r},t)=V({\mathbf r})+gn_0\left [ D-A \left (\frac{x}{X}\right )^2 - 
B \left (\frac{y}{Y}\right )^2 -C \left (\frac{z}{Z}\right )^2 \right ],
\label{eq:mu}
\end{equation}
where $D(t)=1+\epsilon_{dd} F_1$, $A(t)=1+\epsilon_{dd} (F_{1}-F_{2}+F_{3})$, $B(t)=1+\epsilon_{dd} (F_{1}-F_{2}-F_{3})$, 
and $C(t)=1+\epsilon_{dd} (F_{1}+2F_{2})$. Here we have introduced the functions: 
\begin{eqnarray}
\!F_{1}(\chi_x,\chi_y)\!&=&\!\!\int_0^1 \!\!\! du \!\left [ \frac{3u^2}{\sqrt{\alpha^2\!-\!\beta^2}}-1 \right ], \\
\!F_{2}(\chi_x,\chi_y)\!&=&\!\!\int_0^1 \!\!\! du \!\left ( \frac{3u^2\!-\!1}{2} \right )\!\! \left [ \frac{3u^2}{\sqrt{\alpha^2\!-\!\beta^2}}-1 \right ], \\
\!F_{3}(\chi_x,\chi_y)\!&=&\!\!\int_0^1 \!\!\! du \!\left ( \frac{9u^2(1\!-\!u^2)}{4} \!\right )\!\! \left [ \frac{\sqrt{\alpha^2\!-\!\beta^2}-\alpha}{\beta \sqrt{\alpha^2\!-\!\beta^2}} \right ]\!, 
\end{eqnarray}
with $\chi_x(t)=Z(t)/X(t)$, $\chi_y(t)=Z(t)/Y(t)$, $\alpha(\chi_x,\chi_y,u)=(\chi_x^2+\chi_y^2)(1-u^2)/2+u^2$, 
and $\beta(\chi_x,\chi_y,u)=(\chi_x^2-\chi_y^2)(1-u^2)/2$. Substituting Eq.~\eqref{eq:mu} into Eq.~\eqref{eq:SSM1} we obtain a closed 
set of equations for the scaling parameters:
\begin{eqnarray}
 \frac{1}{(2\pi)^2}\frac{\ddot b_x}{\nu_x^2}&=&-b_x + \frac{1}{b_x^2 b_y b_z} \frac{A(t)}{A(0)}, \label{eq:bx}\\
\frac{1}{(2\pi)^2}\frac{\ddot b_y}{\nu_y^2}&=&-b_y + \frac{1}{b_x b_y^2 b_z} \frac{B(t)}{B(0)}, \label{eq:by}\\
\frac{1}{(2\pi)^2}\frac{\ddot b_z}{\nu_z^2}&=&-b_z + \frac{1}{b_x b_y b_z^2} \frac{C(t)}{C(0)}. \label{eq:bz}
\end{eqnarray}
Prior to the quench of $\as$, the stationary TF solution is obtained from the self-consistent equations $\chi_x(0)=\frac{\nu_x}{\nu_z}\sqrt{C(0)/A(0)}$ and 
$\chi_y(0)=\frac{\nu_y}{\nu_z}\sqrt{C(0)/B(0)}$. Solving these equations and using normalisation provides $X(0)$, $Y(0)$, $Z(0)$, and $n(0)$ for known numbers of atoms $N$ and 
trap frequencies $\nu_{x,y,z}$. We use this stationary solution as the initial condition at the start of the quench~($t=0$), and solve the 
system of differential equations~\eqref{eq:bx},~\eqref{eq:by} and~\eqref{eq:bz} to obtain the scaling coefficients. An example for the relevant parameters of Fig.\,3 is shown in  Supplementary Fig.\,3.

\begin{figure}[ht!]
  \centering
\includegraphics[width=0.49\textwidth]{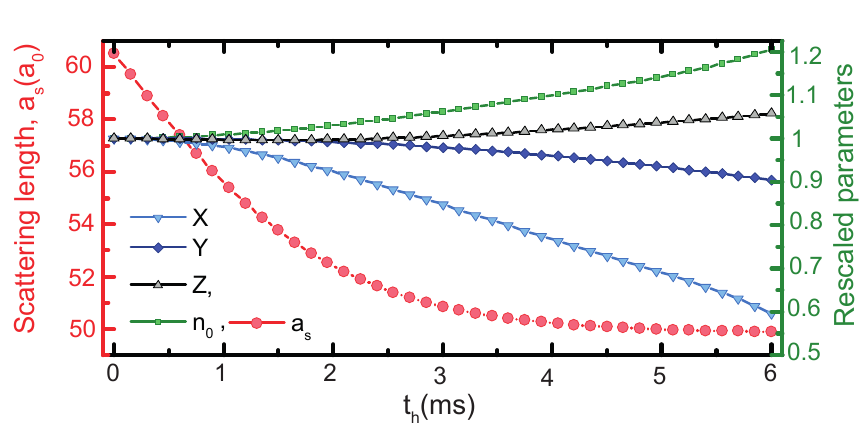} 
\caption{\label{selfsim}
{\bf Self-similar dynamics in quench}. SSM results after a quench from $\asi=61\,a_0$ to $\af=50\,a_0$ in the trap $(\nu_z,\lambda)=(149\,{\rm Hz},4.3)$ and using the experimental cloud characteristics (configuration of Fig.\,3). For reference, we plot the experimentally known $\as(\tho)$ (circles, left axis). We show the 3D TF radii $X$ (down triangles), $Y$ (diamonds), $Z$ (up triangles) and $n_0$ (squares) renormalized by their $\tho=0$-values (right axis). The compression mainly occurs along $X$. The subsequent increase of $n_0$ is less than 2\,\% when $\Delta(y=0)$ touches zero in this configuration (Fig.\,3a). This demonstrates that the dominant effect on the roton spectrum at the instability comes from the reduction of $\as$ itself, even in the case of $\af$ relatively close to $\ar$.}
\end{figure}

\subsection{Numerical simulations of the evolution.}
\label{sec:realtime}

Our simulations of the NLGPE are performed using a split operator  
technique. The evolution operator
over a time $\Delta t$~($\Delta t \to  i \Delta t$ for imaginary time 
evolution)
may be approximately split as
$ e^{-i \hat{H}\Delta t/\hbar}=e^{-i \hat{T}\Delta t/\hbar}e^{-i  
\hat{V}\Delta t/\hbar}+O(\Delta t^2)$. In this expression, $\hat T$ is  
the kinetic energy term,
and $\hat V$ the potential energy. The effective potential energy for  
the evolution is given by the sum of external potential, interparticle interactions,
local LHY correction, and three-body losses (see Section \ref{sec:nlgpe}).

We first evaluate using imaginary time evolution the initial BEC  
wavefunction, $\psi_0(\br)$, prior to the quench of $\as$.
The initial wavefunction for the subsequent real-time evolution is  
then constructed via $\psi_i=\psi_0+\Delta \psi$, where
$\Delta\psi$ accounts for thermal fluctuations, which we simulate by  
populating the excited states of the system as described in Ref.~\cite{Blakie:2008}. The excited states used in this procedure are obtained from the full BdG calculation detailed in Section \ref{sec:bdg}.
Starting with this initial wavefunction $\psi_i$, we mimic as close as  
possible the conditions of our experiments, including ramping,  
holding, and ToF times.
In particular, we include the experimentally calibrated $\as(t)$~(Method \textbf{Quench of the scattering length $\as$}). 
Moreover, for the value of the three-body loss coefficient $L_3$ we use a  
linear fit of the experimentally determined values~\cite{Chomaz:2016}. From the simulated evolution, we obtain the 3D wavefunction of the gas as a function of $\tho$, $\psi(\bf r,\tho)$, from which we can extract the spatial and momentum distributions.

The simulation of the ToF expansion is performed in two steps. First  
we use a multi-grid analysis in order to rescale the size of the  
numerical box
as the cloud expands during the ToF expansion. After some expansion  
time the density drops significantly,
and the subsequent evolution can be readily calculated via $e^{-i  
\hat{T}t/\hbar}$. Our NS
show clearly that the effect of nonlinearity is small during the  
first stages of the evolution, and hence that the ToF expansion indeed
may be employed to image the momentum distribution of the condensate  
at the time in which the trap is opened.

We evaluate the integrated  momentum distribution $\tilde n(k_y,\tho)=\int dk_x dk_z |\tilde \psi({\bf  
k},\tho)|^2$ with $\tilde\psi({\bf k},\tho)$ the Fourier transform of $\psi(\bf r,\tho)$. After $\tho$ of a few ms,  $\tilde n(k_y,\tho)$ shows clear roton peaks. The exact $\tho$ value for the peak emergence depends on the gas characteristics (in particular $T$) and on $\as$.
We evaluate the roton momentum as the mean value of the momentum in  
the roton peak.

From the imaginary-time evolution simulations, we are also able to predict the $\as=\as^{*,\rm st}$ threshold for 
the mean-field instability of the BEC, 
which corresponds to the absence of a mean-field stable solution \cite{Waechtler:2016b,Bisset:2016}. To find this instability boundary, we proceed by steps. We start by calculating the ground-state solution for a given $\as$ that we know to be well within the stable regime. We then reduce $\as$ in small steps and successively calculate the corresponding ground-state solution using the solution of the previous step as starting condition. We do so until no mean-field stable solution
can be found. The predicted $\as^{*,\rm st}$ are reported in Supplementary Table~\ref{table2}. For the theory predictions shown in Fig.~2d (NS and SSM; see Method~\textbf{Analytical dispersion relation for an infinite axially elongated geometry}), we use quenched $\as$ values such that the instability boundary is just slightly crossed, $\as=\as^{*,\rm st}-1a_0$.


\end{document}